\documentclass[manuscript]{aastex}

\slugcomment{To be submitted to the Astrophysical Journal}

\shorttitle{Mid-infrared Emission-Line Diagnostics}
\shortauthors{Weaver et al.}

\begin{document}


\title{Mid-Infrared Properties of the {\it Swift} Burst Alert Telescope Active Galactic Nuclei Sample  of the Local Universe. I.
Emission-Line Diagnostics}


\author{K. A. Weaver, M. Mel\'endez\altaffilmark{1,2}, R. F. Mushotzky\altaffilmark{3}, S. Kraemer\altaffilmark{4},
K. Engle\altaffilmark{5}, E. Malumuth\altaffilmark{6}, J. Tueller, C. Markwardt\altaffilmark{3}}
\affil{NASA Goddard Space Flight Center, Greenbelt, MD, 20771}

\author{C.T. Berghea\altaffilmark{7} and R. P. Dudik}
\affil{U.S. Naval Observatory, Washington, DC 20392}

\author{L. M. Winter\altaffilmark{8}}
\affil{Center for Astrophysics and Space Astronomy, University of Colorado, Boulder, CO 80309-0440 }

\and
\author{L. Armus}
\affil{Spitzer Science Center, California Institute of Technology, Pasadena, CA 91125 }


\altaffiltext{1}{NASA Postdoctoral Program Fellow, Goddard Space Flight Center, Greenbelt, MD, 20771}
\altaffiltext{2}{present address: Department of Physics and Astronomy, Johns Hopkins University, Baltimore, MD, 21218}
\altaffiltext{3}{University of Maryland, College Park, MD 20742}
\altaffiltext{4}{Institute for Astrophysics and Computational Sciences, Department of Physics, The Catholic University of America,Washington, DC 20064}
\altaffiltext{5}{Adnet}
\altaffiltext{6}{SESDA}
\altaffiltext{7}{Computational Physics, Inc., Sprinfield, VA 22151}
\altaffiltext{8}{Hubble Fellow}

\begin{abstract}

We compare mid-infrared emission-line properties, from high-resolution {\it Spitzer} spectra of  
a  hard X-ray (14 -- 195 keV)
selected sample of nearby (z $< 0.05$) AGN detected by the Burst Alert
Telescope (BAT) aboard {\it Swift}. The luminosity distribution for the mid-infrared emission-lines, [O~IV] 25.89 $\mu$m, [Ne~II] 12.81 $\mu$m, [Ne~III] 15.56 $\mu$m and  [Ne~V] 14.32/24.32 $\mu$m, and hard X-ray continuum  
show no differences between Seyfert 1 and Seyfert 2 populations, however 
six newly discovered BAT AGNs are  under-luminous 
in [O~IV], most likely the result of dust extinction in the  host galaxy. 
 The overall tightness of the  mid-infrared correlations and BAT fluxes and luminosities suggests that 
the emission lines primarily arise in gas ionized by the AGN.
We also compare  the mid-infrared emission-lines in the BAT AGNs with those
from published studies of ULIRGs, PG~QSOs, star-forming galaxies and LINERs. We find that the  BAT AGN sample fall into 
a distinctive region when comparing  the [Ne~III]/[Ne~II] and the [O~IV]/[Ne~III] ratios. These line ratios are lower in sources that have been previously classified in the mid-infrared/optical as AGN  than those found for the BAT AGN,  suggesting that, in our X-ray selected sample, the AGN represents  the main contribution to the  observed line  emission. These  ratios  represent a new  emission line diagnostic  for distinguishing  between AGN and star forming galaxies.

\end{abstract}



\section{Introduction}

Active galactic nuclei (AGN) span over seven orders
of magnitude in bolometric luminosity ($L_{bol}$) \citep{1999PASP..111....1K} and yet
are all believed to be powered by the 
same physical mechanism: accretion of matter onto supermassive black holes \citep[e.g.,][]{1984ARA&A..22..471R,2004ApJ...613..682P}. One way to approach the study of AGN is to concentrate on those in the local Universe (e.g. $z < 0.05$), which permits us, among other things, to determine the properties
of the host galaxy. Such studies tend to focus on Seyfert galaxies, which are
modest luminosity AGN ($L_{bol} \lesssim 10^{45}$ erg s$^{-1}$), but
bright enough, due to their proximity, to be studied across the full
electromagnetic spectrum.

Although Seyfert galaxies and other AGN have been traditionally defined in terms of their optical properties \citep[e.g., classification into  Type I and Type II,][]{1974ApJ...192..581K},
sample selection of AGN via a single waveband can lead to observational 
bias \citep[e.g.,][]{1994ApJ...436..586M}. For example, most AGN are obscured from our line 
of sight by dust and gas \citep{2000A&A...355L..31M} and any selection based
on optical (or UV) properties would miss many objects or could highly skew a sample
towards unobscured objects \citep[e.g.,][]{2005AJ....129..578B}. The soft X-ray properties 
of Seyfert galaxies generally follow the same dichotomy as
their optical properties. The X-ray continuum source in Seyfert~1s can
be observed directly  \citep[e.g.,][]{1998ApJS..114...73G},
while the central X-ray source is sometimes undetectable in Seyfert 2s, due to material 
 with ${\rm N_H} >$ 10$^{22.5}$~cm$^{-2}$ along our line of sight. There is additionally a wide range of effective IR colors \citep{2003ApJ...590..128K,2005ApJ...632L..13L} 
which can introduce selection bias. A comparison of infrared and X-ray data \citep{2005AJ....129.2074F} 
shows a factor of 30 range in the IR 24~$\micron$ to $\sim$4 keV X-ray flux ratio for X-ray selected AGN, suggesting a range 
of geometries and optical depths for dust reprocessing, and probably variance in the 
intrinsic power law AGN continuum. Obscuration and star formation in the host galaxy 
can  also  dominate and introduce confusion   in the IR  \citep[e.g.,][]{2004A&A...418..465L,2006ApJ...642..126B}. 
In fact, virtually all surveys for AGN based purely on IR, optical, UV or soft X-ray data have been biased
\citep{2004ASSL..308...53M}.  Even Sloan surveys \citep{2004ApJ...613..109H} or 
 IR surveys \citep{2005AJ....129.2074F} have required indirect AGN indicators which are known to be not necessarily robust \citep{2008ApJ...682...94M}.

To understand the intrinsic properties of AGN as a class, it is critical to start with a
survey where we can be as certain as possible that we are viewing the AGN-only parts of these galaxies. 
At X-ray energies of E $> 10-20$ keV, the obscuring material is relatively optically thin for column densities less 
than $\sim 3 \times 10^{24}$ cm$^{-2}$ (Compton-thin objects). Even if an 
AGN is well buried within its host galaxy there is an unaffected view of the central power source.
A hard X-ray survey should thus find all Compton thin AGN in a uniform fashion and 
is the most representative, since at present, there are very few, if any, known X-ray ``quiet" AGN.
Such a hard X-ray survey is now available from the {\it Swift} Burst Alert 
Telescope (BAT). The Swift BAT is sensitive over $\sim$85\% of the 
sky to a flux threshold of 
$2 \times 10^{-11}$ ${\rm ergs~cm^{-2} s^{-1}}$ in the 14$-$195~keV band \citep{2005ApJ...633L..77M}. 
The BAT data are about 10 times more sensitive than the previous hard 
X-ray all sky survey \citep{1984ApJS...54..581L}. The BAT detects all bright AGN, whether they are 
obscured or not. Moreover,  several of the BAT sources are newly discovered AGN or have 
been poorly studied, if at all, at other 
wavelengths \citep{2008ApJ...674..686W,2008ApJ...681..113T,2010ApJS..186..378T}.

Nevertheless, although all of the BAT-detected objects are true AGN, in order to fully explore
the properties of these AGN one needs to take a multi-wavelength approach. For example,
studying the IR properties of the BAT AGN will provide insight into the IR/X-ray scatter 
and thus determine the true distribution of IR properties. There have been a large number of studies of the mid-infrared emission line properties of active galaxies using both {\it Infrared Space Observatory} \citep{1996A&A...315L..27K} and {\it Spitzer Space Telescope} \citep{2004ApJS..154....1W}. The ratios of high- and low-ionization mid-infrared emission lines have been widely used to separate the relative contribution of the  AGN and star formation \citep[e.g.,][]{1998ApJ...498..579G,2002A&A...393..821S,2006ApJ...646..161D,2007ApJ...656..148A,2007ApJ...667..149F,2008ApJ...689...95M,2010ApJ...710..289B}. More recently, \cite{2009ApJ...704.1159H} (H09) used new high-resolution {\it Spitzer} spectroscopy to probe the utility of mid-infrared emission line diagnostics as a way to separate active galaxies from star forming galaxies. 
In our first  study  of mid-infrared properties of the BAT AGNs \citep{2008ApJ...682...94M}, we found the [O~IV]25.89$\mu$m to be an accurate indicator of the AGN luminosity, with an uncertainty of $\sim$0.3~dex; this result has been   confirmed  using   larger samples  \citep{2009ApJ...700.1878R,2009ApJ...698..623D}. Using a complete, volume-limited,   sample of galaxies \cite{2009MNRAS.398.1165G}  (GA09) demonstrated the utility of high-ionization mid-infrared emission lines, such as [Ne~V]14.32$\mu$m,  to identify AGN including those that were not identified as AGN in optical studies \citep[see also,][]{2007ApJ...669..109L,2008ApJ...678..686A,2009ApJ...691.1501D,2009ApJ...704..439S}. Similar results have been found by \cite{2009ApJS..184..230B} (B09) in their study of starburst galaxies.

This paper is the first in a series seeking to understand the nature of the observed mid-infrared  
luminosities in AGN and their wide variety of spectral forms.
Here we report results from the portion of our sample that have
high-resolution {\it Spitzer} spectra. This work complements the extensive optical imaging and spectroscopy of the AGN population \citep[Koss, accepted in ApJL;][]{2010ApJ...710..503W} and the detailed analysis of the X-ray properties of the 
BAT AGN sample \citep[e.g.,][]{2008ApJ...674..686W,2009ApJ...690.1322W,2009ApJ...701.1644W}. In following papers we will report on the results from our analysis of the low-resolution {\it Spitzer} spectra which will include the study of polycyclic aromatic hydrocarbon (PAH) features, silicate absorption and mid-infrared continuum properties of the BAT AGN  sample. In order to calculate the luminosities presented in this work  we assumed a flat universe with a Hubble constant $H_o=71{\rm kms^{-1}Mpc^{-1}}$, $\Omega_\Lambda=0.73$ and ${\rm \Omega_M=0.27}$,
with redshift values taken from NASA's ExtraGalactic Database (NED), except for sources with redshift values of $z<0.01$ were distances are take from The Extragalactic Distance Database (EDD) \citep{1988ngc..book.....T,2009AJ....138..323T}.  

\section{Observations and Analysis}

\subsection{X-ray data and AGN sample selection}

Our sample was selected from the first unbiased local AGN sample obtained with
the Swift Burst Alert Telescope (BAT) survey.    
The X-ray positions have $\sim3^{''}$  positional uncertainties and we are highly confident of the 
positions and the optical identifications. Source identifications are based 
primarily on the X-ray imaging data and a correlation with optical images and catalogs. In some
cases the identifications are based on positional coincidences with previously known AGN.  

The median redshift of the BAT objects is $z \sim$ 0.025. Our selection criteria are $z< 0.05$,
$|b| > 19^{\circ}$ and a hard X-ray BAT flux of  $ > 2 \times 10^{-11}$~erg~cm$^{-2}$~s$^{-1}$. The flux limit provides 
sufficient S/N to be sure that the sources are statistically robust. 
Our total BAT sample contains 130 objects above a significance threshold of 5.0 $\sigma$. We obtained  from the Infrared Spectrograph (IRS) \citep{2004ApJS..154...18H} on board {\it Spitzer}   high and low resolution 
spectra for sixty of these objects, while seventy previously had {\it Spitzer} IRS spectra from 
other observing programs. There are published BAT AGN 
catalogs from two surveys, the 9-month survey \citep{2008ApJ...681..113T} and an 
updated 22-month catalog \citep{2010ApJS..186..378T}. Here we use the 
22-month fluxes from \cite{2010ApJS..186..378T}  as the 22-month survey is the most sensitive. However, in our sample,  four of the AGN in the 9-month survey were not detected in the 22-month survey, therefore we will use the fluxes from the 9-month survey as upper-limits (see footnote in Table~1).



\subsection{{\it Spitzer} Observations and Data Analysis}

In  this work we discuss only the high-resolution  {\it Spitzer} spectroscopy of our total sample, which result  on a subsample of 79 BAT AGN. The galaxies in 
our sample were observed with IRS  in  the 
Short-High (SH, $\lambda$ = 9.9 - 19.6 $\mu$m, 4.7$'$$'$ $\times$ 11.3$'$$'$,$R\sim 600$) 
and Long-High (LH, $\lambda$ = 18.7 - 37.2 $\mu$m, 11.1$'$$'$ $\times$ 22.3$'$$'$, 
$R\sim600$)  IRS order   in  staring mode. Science targets in staring mode are placed at two node 
position along the IRS slit. This sample includes fluxes found in the literature \citep{2005ApJ...633..706W,2006ApJ...640..204A,2008ApJ...676..836T} and from  our analysis 
of unpublished archival spectra observed with IRS, including observations collected with 
the {\it Spitzer} Cycle 3 and Cycle 5 GTO program ``Spitzer Observations to Complete the 
First Unbiased AGN Sample of the Local Universe" (P30745 and P50588, PI: K. Weaver). 
This high-resolution sample includes  38 Seyfert 1's, 33 Seyfert 2's, 6 previously unknown or poorly-studied   
AGNs and two  LINERs. High resolution spectroscopy allow us to easily separate blended features such as,  [O~IV]-[Fe~II]~25.99$\mu$m and  [Ne~V]~14.32$\mu$m-[Cl~II]~14.37$\mu$m. Moreover, the subsample presented in this work spans the same range on 14-195~keV  luminosities as the 22-month BAT AGN sample and is large enough to be statistically representative of the whole sample of 130 AGN. Furthermore, the Kolmogorov-Smirnov (K-S) test probabilities for Seyfert~1 and Seyfert~2 galaxies between the subsample presented in this work and the whole sample of 130 sources are, $46.2\%$ and $11.5\%$\footnote{A probability  value of less than 5$\%$ represents a high level of significance that two samples drawn from the same parent population would differ this much 5$\%$ of the time, i.e., that they are different. A strong level of significance is obtained for values smaller than 1$\%$ \citep[e.g.,][]{1992nrfa.book.....P,2003drea.book.....B}}, respectively, meaning that there are no differences in their X-ray luminosities. Finally, our  sample also has  the same range in mid-infrared emission line luminosities  as   complete samples such as the 12$\micron$ \citep{2010ApJ...709.1257T} and the revised Shapley-Ames sample \citep{2009ApJ...700.1878R}.

For the analysis of  {\it Spitzer} data we used the basic calibrated data files preprocessed 
using the S17.2 IRS pipeline. This include ramp fitting, dark sky subtraction, drop correction, linearity 
correction and wavelength and flux calibrations \footnote{See the IRS Pipeline Handbook, 
http://ssc.spitzer.caltech.edu/irs/dh/irsPDDmar30.pdf}. Many of the sources in our sample have 
dedicated off-source observations to do  sky subtraction, which can alleviate the effect of rogue 
pixels and variable background. Off-source images were averaged for each node position and order 
to obtain a final background image. Then, we subtract the sky background from our spectrum. The full 
slit  spectra were extracted from the IRS data  using the Spectroscopy Modeling Analysis and Reduction 
Tool (SMART) v6.4.0 \citep{2004PASP..116..975H}.  For the extraction we used the ``Full Aperture" extraction 
method for  high-resolution observations of point sources. We created median
basic calibrated data files from each node and 
then the spectra from each node position for SH/LH  were averaged using 2.5-$\sigma$ clipping to reject 
outliers.  Finally, we trimmed the edges of the orders to obtain a  clean spectrum. For sources without 
dedicated off-source  observations we did not perform any background subtraction because  we only 
required emission-line fluxes, furthermore,  our hard X-ray selected sample is characterized  by bright 
nuclear sources which fill the high-resolution slit resulting in a minimal background correction. We 
performed the line fit with SMART using a polynomial to fit the  continuum  and a Gaussian for the 
line profile. In Table~1 we  report the line fluxes, together with their 1-$\sigma$ statistical error for the whole high-resolution sample. The typical 1-$\sigma$ statistical errors for the mission lines presented in this work are, on average, $\sim$9$\%$, $\sim$10$\%$, $\sim$5$\%$, $\sim$12$\%$ and $\sim$8$\%$ for the [Ne~II]12.81$\mu$m, [Ne~V]14.32$\mu$m, [Ne~III]15.56$\mu$m, [Ne~V]24.32$\mu$m and [O~IV] emission lines, respectively.  From this and for the sake of simplicity  we  did not plot error bars on individual objects in  the different comparisons presented in this work, as the errors are comparable to the symbol size uses in the figures. For non-detections we quote the 3-$\sigma$ upper limits 
as defined for  emission-lines with a S/N less than 3. Finally, the emission line fluxes are  presented here without reddening corrections.

 Each of the galaxies in our sample has clearly detected
[O~IV], [Ne~II] and [Ne~III] emission. However, [Ne~V]~14.32$\mu$m and 24.32$\mu$m  were not detected in $\sim$ 10$\%$ and $\sim$ 15$\%$ of the sources, respectively. Similar results are obtained for the [Ne~V]~14.32$\mu$m and 24.32$\mu$m  in the {\it Spitzer} high-resolution spectroscopy of the 12$\mu$m Seyfert galaxies \citep{2008ApJ...676..836T,2010ApJ...709.1257T}. The importance of  [O~IV] and the neon emission lines is that they are  sufficient to 
distinguish  between stellar and AGN activity. For example, the [Ne~III]/[Ne~II] and [O~IV]/[Ne~II] ratios
are good discriminators of star formation and  AGN emission \citep{1998ApJ...498..579G,2002A&A...393..821S,2004A&A...414..825S,2008ApJ...689...95M}, while 
the strengths of [O~IV] \citep{2008ApJ...682...94M,2009ApJ...700.1878R} and [Ne~V] \citep[e.g.,][]{2007ApJ...663L...9S,2007ApJ...664...71D,2008ApJ...678..686A} scale with the  luminosity of the AGN. Moreover, the neon line ratios are also tracers that are insensitive to abundance. It should be noted,
however, that [O~IV] and [Ne~V] can be weak  in some classical AGN \citep{2005ApJ...633..706W}, notably ultraluminous infrared galaxies,  possibly due to the shielding of the narrow line region by optically thick, dusty gas close to the AGN \citep[e.g.,][]{2007ApJ...656..148A}.

\section{Mid-Infrared properties of the BAT sample}

\subsection{Comparison of the IR Emission-Line and BAT Luminosities}

\begin{figure}
\epsscale{0.9}
\plotone{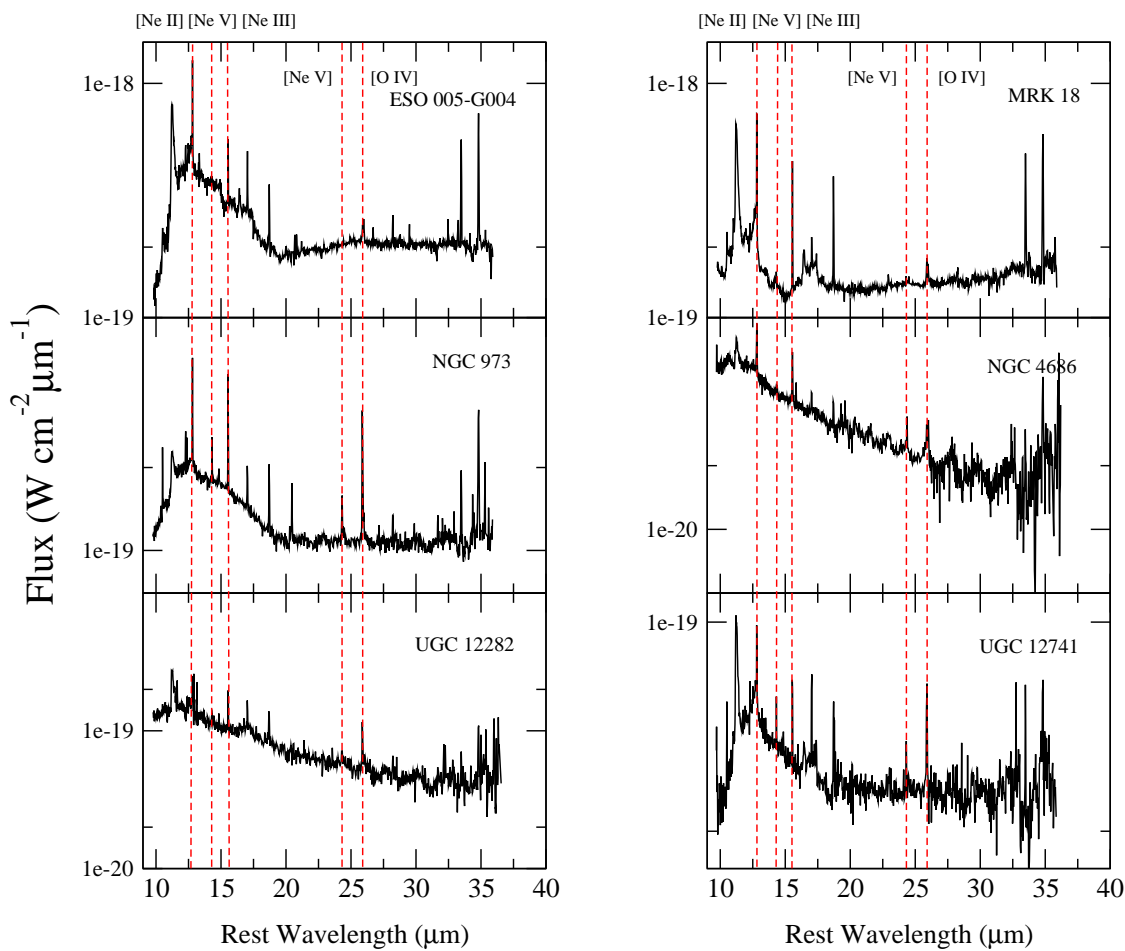}
\caption{Combined {\it Spitzer} IRS SH and LH spectra for the newly detected AGN
in the BAT sample. The [Ne~II] 12.81 $\mu$m, [Ne~III] 15.56 $\mu$m, [Ne ~V] 14.32 $\mu$m, [Ne~V] 24.32 $\mu$m,
and [O~IV] 25.89 $\mu$m lines are indicated by the vertical lines. The strong emission line features at $\sim$33~$\mu$m and $\sim$35~$\mu$m correspond to [S~III] and [Si~II], respectively.\label{fig1} }
\end{figure}

As noted above, via the hard X-ray band  we can detect AGN that may have been missed in optical surveys due
to the extinction of their optical emission lines. Among the 79 targets present here (BAT AGN), we have found
six poorly studied or previously unknown AGN (``new BAT-detected AGN"): ESO~005-G004, MRK~18, NGC~973, NGC~4686,  UGC~12282, 
and UGC~12741. There have been  attempts to classify some of these galaxies. Of these,
ESO~005-G004 has been classified as a Seyfert 2, based on its X-ray properties  although no  optical AGN signature has been detected \citep{2007ApJ...669..109L,2007ApJ...664L..79U}. Based on follow-up observations in the optical, MRK~18 has an ambiguous classification  with HII/LINER properties \citep{2010ApJ...710..503W} and NGC~973 has been classified as a narrow emission line AGN with Sy2/LINER properties \citep{2008A&A...482..113M}. UGC~12282 has been classified as a Sy 1.9 based on its optical spectra \citep{2006A&A...455..773V}. Finally, to the best of our knowledge, there is no assigned classification for NGC~4686 and  UGC~12741.  The {\it Spitzer} IRS spectra of these objects are shown in Figure~\ref{fig1}.

In Figures~2 -- 7 we compare the mid-infrared emission lines and the BAT luminosities (and fluxes),
correlations between the emission lines, and emission line ratios.
The  statistical analysis for these plots is listed in Table~2, which 
includes the Spearman rank order correlation coefficient with its associated null probability, 
the generalized Kendall's correlation coefficient for censored data and the Kendall's  
coefficient for partial correlation with censored data.  One should note that, 
due to redshift effects, luminosity-luminosity plots will almost always show some correlation. Thus, we are primarily interested in the dispersion of the correlations or the slopes.  
Furthermore, caution must be taken when applying statistical analysis to data sets that contain 
non-detections (upper limits), or  ``censored'' data points. To deal with these problems we 
have used  Astronomy SURVival analysis ({\bf ASURV}) Rev 1.2  \citep{1990BAAS...22..917I}, which implement the methods presented in  \cite{1986ApJ...306..490I}. We also used a test for partial 
correlation with censored data \citep{1996MNRAS.278..919A} in order to exclude  the redshift effect in the correlations.

\begin{figure}
\epsscale{0.9}
\plotone{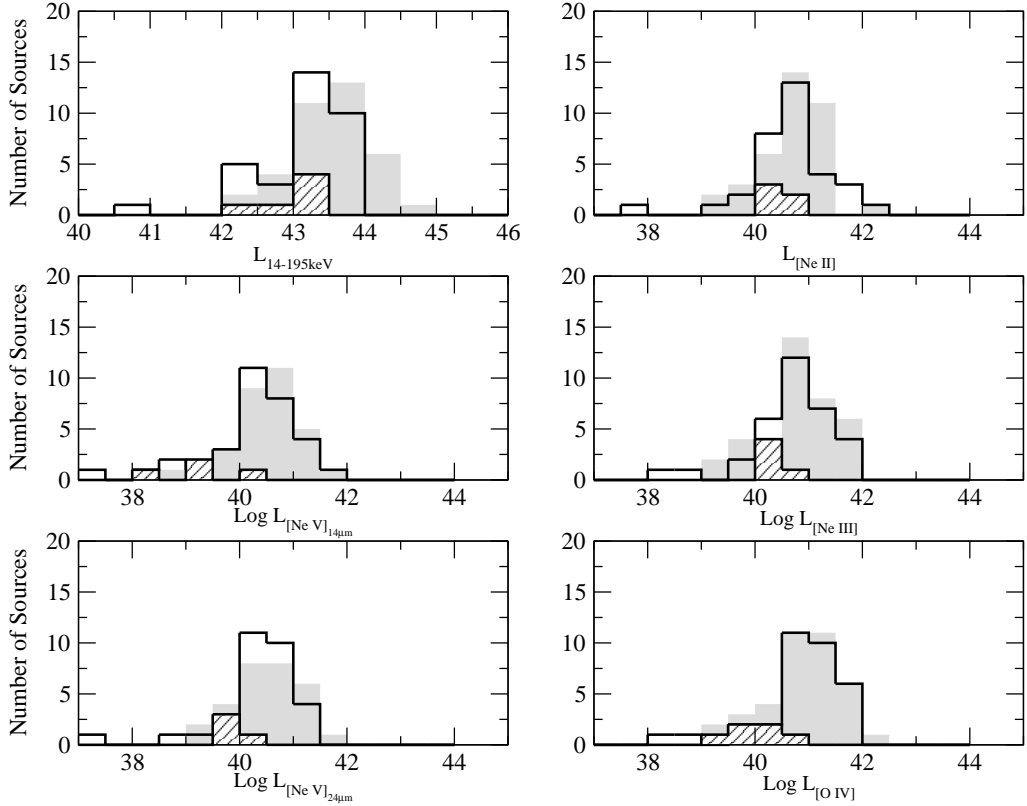}
\caption{Histograms for the 
mid-infrared and BAT  (14- 195 keV) luminosities for our sample. Seyfert~1 and Seyfert~2 galaxies are indicated with gray bins and solid lines, respectively. The newly detected BAT AGN are indicated in bins with dashed lines; note
 that their [Ne~III] and [O~IV] luminosities appear to be low, i.e., compared to AGN with similar BAT 
luminosities. The K-S test for these emission line luminosities show that two samples drawn from the same population would differ this much $\sim 6.3\%$, $\sim 82.6\%$, $\sim 69.8\%$, $\sim 50.9\%$, $\sim 28.6\%$ and $\sim 98.1\%$  of the time for the BAT, [Ne~II], [Ne~V]~14$\mu$m, [Ne~III], [Ne~V]~24$\mu$m and [O~IV] luminosity distributions, respectively.\label{fig2}}
\end{figure}

In Figure~\ref{fig2} we show the distribution of the observed [Ne~II], [Ne~III], [Ne~V], [O~IV] and BAT luminosities ($\rm{ L_{BAT}}$)
for our sample. The results of the K-S tests, comparing Seyfert 1s and 2s,
for these quantities are listed in Table~3. This table also includes 
information about the numbers of Seyfert 1 and Seyfert 2 galaxies, median values and standard deviations of the mean for the measured quantities. As is apparent from both the histograms and the K-S results,
the two Seyfert types are statistically indistinguishable, i.e., there are
essentially no differences between Seyfert 1s and 2s when we directly compare the BAT 
luminosities or any of the IR line strengths. These results are in agreement with the similar optical luminosities found for broad and narrow line sources, with optical Seyfert classifications, in the most recent study on the optical spectral properties of the BAT AGN \citep{2010ApJ...710..503W}. 
The smallest null-test probability is for ${\rm L_{BAT}}$, which, as we will discuss below, is evidence that
 Seyfert 2s suffer more from Compton scattering  in the BAT band. Although there were too few
of the ``new BAT AGN" to separate them for a K-S on these quantities, the
histograms indicate that they are relatively weaker in their observed mid-infrared emission line luminosities,
compared to their BAT luminosities.


As mentioned before, a detailed statistical analysis for the different correlations between the BAT and mid-infrared luminosities is presented in Table~2. From this analysis the weaker correlation found in the [Ne~II] - BAT relationship 
could be the result of active star formation contributing to the [Ne~II] emission line in some of the AGN \citep[e.g.,][]{2002A&A...393..821S,2006ApJ...649...79S,2007ApJ...658..314H,2008ApJ...689...95M}. On the other hand, stronger  correlations are found between  [Ne~III], [Ne~V] and [O~IV] luminosities  when compared to ${\rm L_{BAT}}$, suggesting that, on average,  there is no strong enhancement due to  star formation  in the [Ne~III] and [O~IV]  emission in the BAT sample.

\begin{figure}
\epsscale{0.9}
\plotone{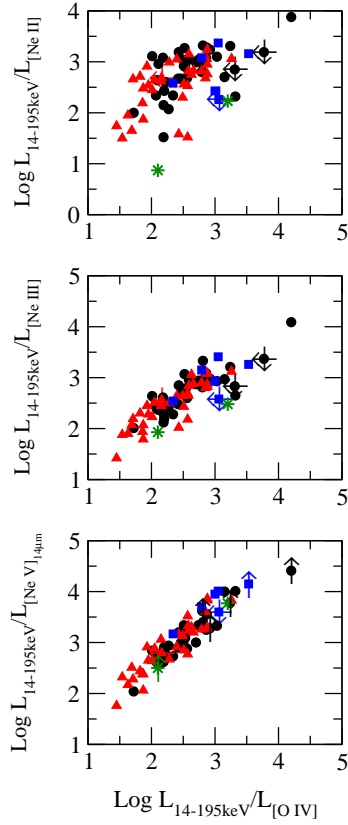}
\caption{Comparison between the ratios of the BAT band and the mid-infrared 
emission lines. Seyfert~1 galaxies are presented as black circles, Seyfert~2 galaxies are red triangles, blue squares represent the newly detected BAT AGN and green stars are LINERs. Of particular interest 
is the statistically significant (see Table 3) separation 
between Seyfert types in 
the ${ \rm L_{14-195keV}/L_{[O~IV]}}$ ratio, in agreement with this 
ratio being an indicator of Compton scattering in the BAT band. One must note  that  below $ {\rm L_{BAT}/[O~IV]< \sim 2.0} $ there are  only two Seyfert 1 galaxies but 
 thirteen  Seyfert 2 galaxies ($\sim 40 \%$ of the Seyfert 2 population)\label{fig4}}
\end{figure}

\begin{figure}
\epsscale{0.9}
\plotone{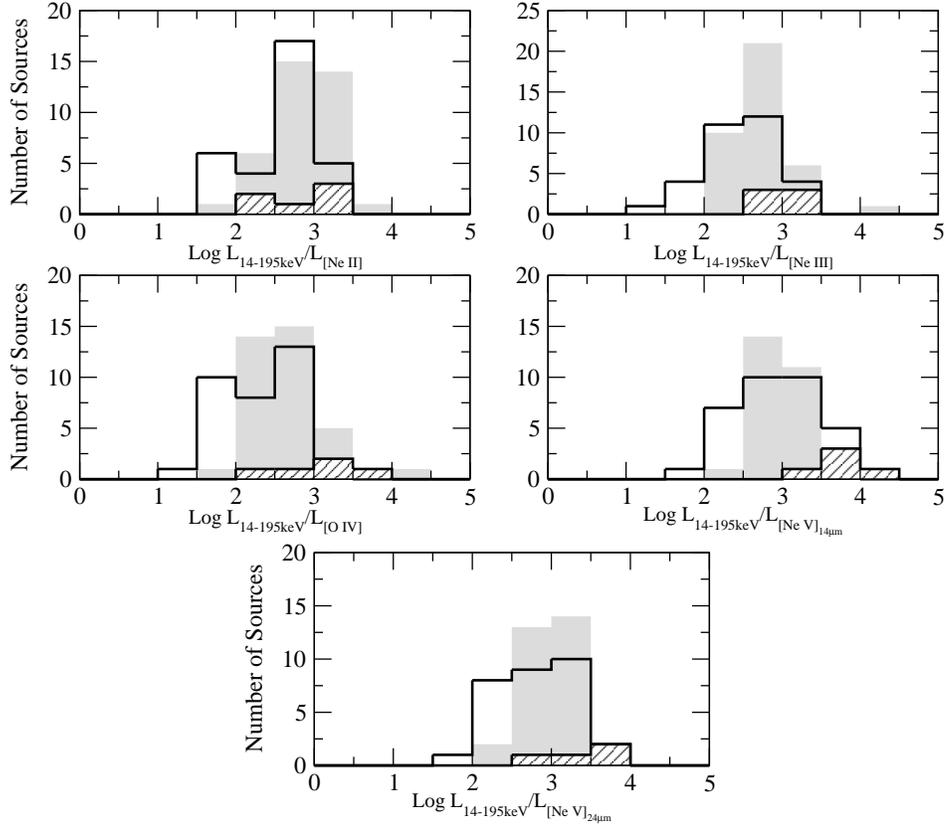}
\caption{Histograms for the ratios of the BAT band and the mid-infrared 
emission lines. Seyfert~1 and Seyfert~2 galaxies are indicated with gray bins and solid lines, respectively. The newly detected BAT AGN are indicated in bins with dashed lines.  Interestingly, all these ratios are log normal distributed with a standard deviation of only 0.5~dex, except for the ${ \rm L_{14-195keV}/L_{[Ne III]}}$ with a standard deviation of 0.4~dex (see Table~3). The K-S test for these ratios show that two samples drawn from the same population would differ this much $\sim 2.8\%$, $\sim 9.0\%$, $\sim 2.4\%$, $\sim 26.1\%$ and $\sim 9.5\%$ of the time for the ${ \rm L_{14-195keV}/L_{[Ne~II]}}$, ${ \rm L_{14-195keV}/L_{[Ne~III]}}$, ${ \rm L_{14-195keV}/L_{[O~IV]}}$, ${ \rm L_{14-195keV}/L_{[Ne~V](14.32\mu m)}}$ and ${ \rm L_{14-195keV}/L_{[Ne~V](24.32\mu m)}}$ distributions, respectively.\label{fig5}}
\end{figure}

In Table~4 we present  the linear regression fits for all  correlations. The somewhat steeper slopes for the Seyfert~2s are consistent with the decrement in the observed X-ray emission in the BAT band  due to the effect of  Compton scatter. The strong similarities between Seyfert types lessen when we compare the {\it ratios} 
of mid-infrared line strengths and the X-ray continuum strength. In Figure~\ref{fig4}, we show the ratios of ${\rm L_{BAT}}$ and [Ne~II], [Ne~III], and [Ne~V]~14.32 $\mu$m compared
to ${\rm L_{BAT}}$/[O~IV]. The histograms for the ratios   of ${\rm L_{BAT}}$ and [Ne~II], [Ne~III], [Ne~V]~14.32/24.32 $\mu$m and [O~IV] are presented  in Figure~\ref{fig5}. Of particular interest in these 
comparisons is  that  below $ {\rm \log L_{BAT}/[O~IV]< \sim 2.0} $ there are  only two Seyfert 1 galaxies but 
 thirteen  Seyfert 2 galaxies ($\sim 40 \%$ of the Seyfert 2 population in our sample). Therefore, in these plots, one can see that the Seyfert~2s have lower ratios of ${\rm L_{BAT}}$ to the emission lines compared to the Seyfert~1s, which is, we believe  is due to  the effect of   Compton scattering in the BAT band in some Seyfert 2 galaxies. This result is in agreement with the K-S test null probability   for this ratio (see Table~3). Similarly, there is only one Seyfert~2 galaxy, NGC~4138, with $ {\rm \log L_{BAT}/[O~IV]> \sim 3.0}$.

  From the ${\rm L_{BAT}/[O~IV]}$ ratios, the lowest values, likely due to   the effects of Compton scattering in the BAT band are: NGC~1365, NGC~2992, NGC~3281, NGC~7582, MRK~3 and MCG-03-34-064. Note that NGC~2992 and NGC~1365 are very variable in the X-ray, which can contribute to their low ${\rm L_{BAT}/[O~IV]}$ and   ${\rm L_{BAT}/[Ne~V]}$ ratios \citep[see,][]{2000A&A...355..485G,2000A&A...356...33R}. In the  BAT spectra  (Mushotzky et al. 2010, in preparation), MRK~3 shows a flat spectrum with a  photoelectric cutoff, consistent with   a high column density, but   not a Compton thick absorber. On the other hand, NGC~1365, NGC~3281 and NGC~7582 are most probably Compton thick sources and the BAT data for NGC~2992 and MCG-03-34-064 has  a low S/N which makes it difficult to constrain the spectrum well. These results suggest that the low  BAT  to [O~IV] ratio is a marker for very high column densities towards the X-ray source, in agreement with previous studies \citep{2008ApJ...682...94M,2009ApJ...700.1878R,2009ApJ...705..568L}.  These results argue in favor of the unified model of active galaxies, where the amount of X-ray emission suppressed, relative to an  isotropic indicator for the AGN power, is related to the absorbing column density towards the X-ray source \citep[see also][]{1999ApJS..121..473B,2005ApJ...634..161H,2006A&A...453..525N}. Given the high ionization potential for Ne$^{+4}$ ($\sim$97~eV), the mid-infrared [Ne~V] emission lines are claimed to be an unambiguous signatures of an AGN \citep[e.g.,][]{2007ApJ...663L...9S,2009ApJ...704..439S}. Therefore,  the very tight correlation between the ${\rm L_{BAT}/[O~IV]}$  and ${\rm L_{BAT}/[Ne~V]}$~14.32$\mu$m  ratio  confirm  that [O~IV] is AGN-driven in this sample.

\begin{figure}
\epsscale{0.9}
\plotone{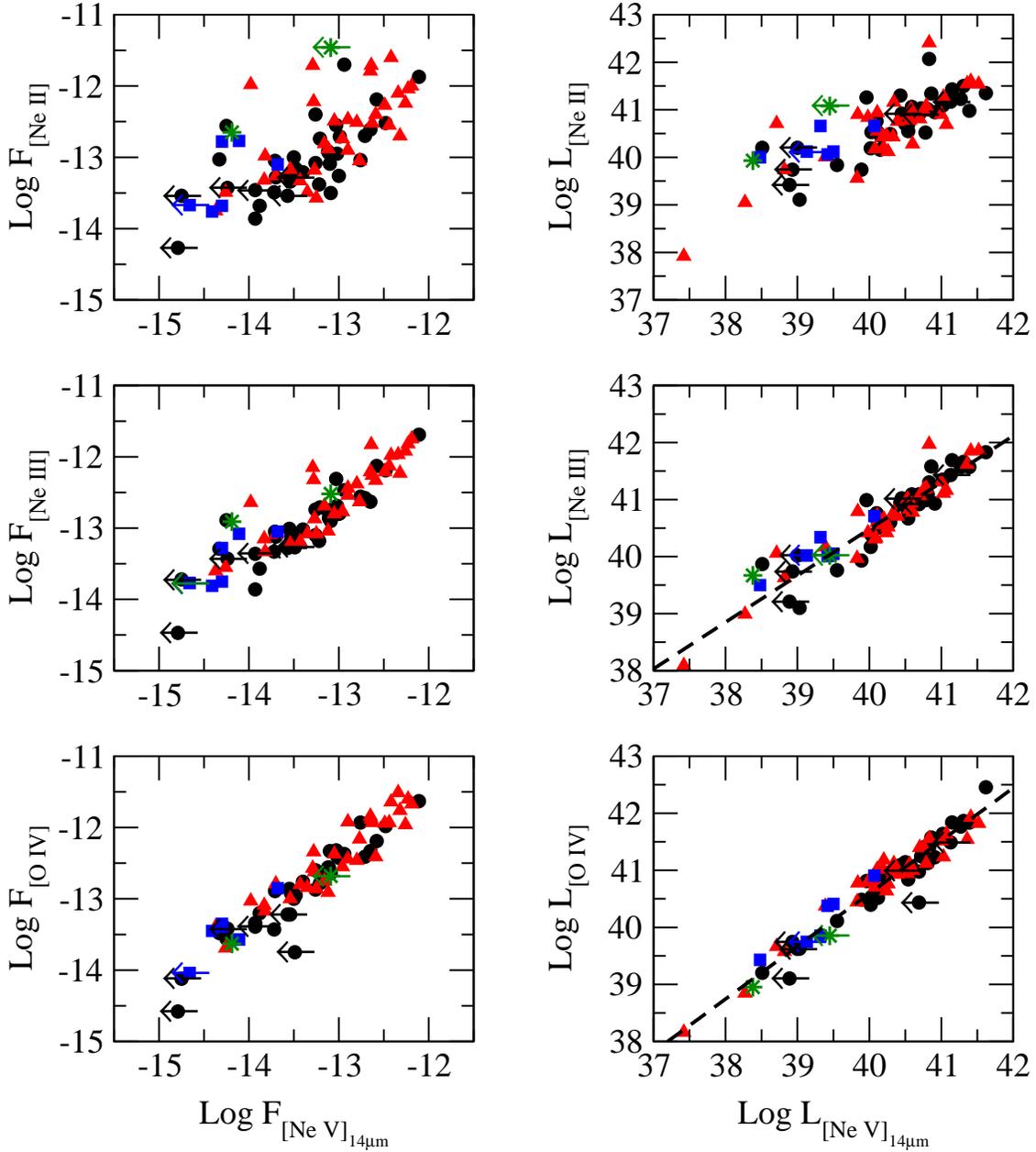}
\caption{Correlation between [\ion{Ne}{2}],[\ion{Ne}{3}], 
[\ion{Ne}{5}]$14\mu m$ and [\ion{O}{4}] fluxes and luminosities. 
Symbols are the same as in Figure~\ref{fig4}. The dashed line represents the 
linear regression for the full sample (see Table~4). The statistical analysis and linear regression fits  for these correlation are presented in  Table~2 and Table~4, respectively. \label{fig6}}
\end{figure}

In Figure~\ref{fig6} we plot both the flux and luminosity of [Ne~II], [Ne~III] and [O~IV] against those of [Ne~V]~14.32$\mu$m. 
The very tight correlation between [Ne~V] and [O~IV] suggest 
that both of these lines are produced by the same physical process, i.e. photoionization by the AGN continuum \citep[see also,][]{2008ApJ...682...94M,2008ApJ...689...95M}. This is also true for the tight correlation between [Ne~V] and [Ne~III]  found  in our sample.  While there is  some scatter for [Ne~III], which may also be due to star formation \citep{2007ApJ...658..314H}, the tightness of the correlation suggests that [Ne~III] is primarily produced by the AGN in these objects \citep[see also][]{2007ApJ...655L..73G,2009MNRAS.398.1165G}, in agreement with the good correlation found between the [Ne~III] and BAT luminosity. Furthermore, the tightness of these  correlations, specially in flux-flux, suggests that the constant mid-infrared ratios observed are not dominated by aperture effects. The extraction aperture for the [O~IV] is bigger than that for the [Ne~III] and [Ne~V]14$\mu$m, implying that the emitting regions  where these lines originate are well within the central kpc of these sources,  in agreement with our previous photoionization studies \citep[e.g.,][]{2008ApJ...682...94M,2008ApJ...689...95M}. We found that none  of the correlations presented, e.g., between the mid-infrared lines and the BAT luminosities, are  driven by distances effects, as shown by the partial correlation analysis (see Table~2) and in agreement with the tightness of the flux--flux correlations. Finally,  the largest scatter occurs for [Ne~II], which is consistent with the interpretation that this line may contains a significant contribution from stellar processes, resulting in the enhancement of this emission line, as also evident in Figure~\ref{fig4}. As we discussed before, the  [Ne~V] emission lines are good indicators for the AGN power, therefore, the good correlation found  between [Ne~II] and [Ne~V] (see Table~2) suggest  that, on average, the observed [Ne~II] emission in the BAT sample is still dominated by the AGN.

\begin{figure}
\epsscale{0.9}
\plotone{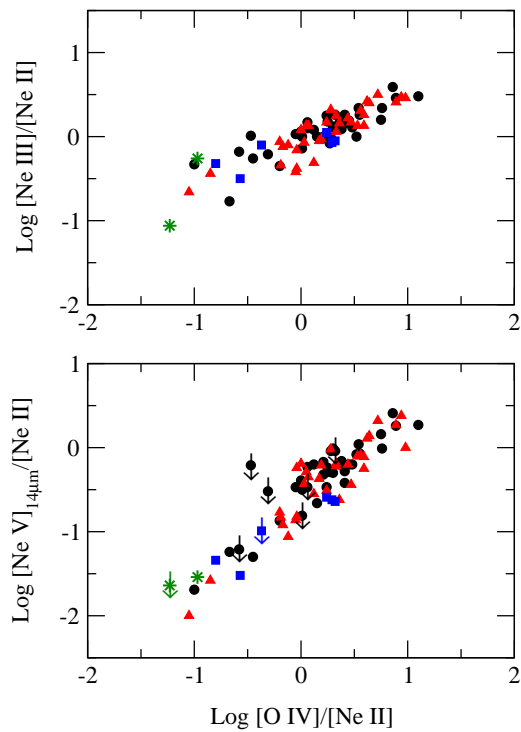}
\caption{[Ne~III]/[Ne~II] and  
[Ne~V]$14\mu m$/[Ne~II] versus [O~IV]/[Ne~II] ratios for Seyfert 1, Seyfert 2, 
BAT-detected AGNs and LINERs in the BAT sample. Symbols are the same as in Figure~\ref{fig4}. The $\sim 62\%$ and $\sim 72\%$ of the objects in our sample  have
[Ne~III]/[Ne~II] and [O~IV]/[Ne~II] have ratios bigger than unity, which has been shown to be typical of
AGN-ionized emission-line gas \citep[e.g.,][]{2008ApJ...689...95M}. The 
the [Ne~V] and [O~IV] ratios appear well-correlated, as expected given the
tight correlation seen in Figure~\ref{fig6}.\label{fig7}}
\end{figure}

As we mentioned before, the ratios of high-ionization lines to low-ionization lines can reveal
the relative contributions of the AGN and star formation. In Figure~\ref{fig7}, we show the ratios of [Ne~III]/[Ne~II]
and [Ne~V]14$\mu$m/[Ne~II] compared to [O~IV]/[Ne~II] for the BAT sample. One result, which was also
discussed in \cite{2008ApJ...689...95M},  is that $\sim 62\%$ and $\sim 72\%$ of 
our AGN have [Ne~III]/[Ne~II] and [O~IV]/[Ne~II] ratios, respectively,  greater than unity. This suggests, again,  that in the BAT targets, the mid-infrared emission lines are dominated by the AGN. These results are also in agreement with the high-resolution {\it Spitzer} analysis of the 12$\mu$m sample  by \cite{2008ApJ...676..836T,2010ApJ...709.1257T} where they found the same range for the [Ne~III]/[Ne~II] and [O~IV]/[Ne~II] ratios. However, the K-S test for the [Ne~III]/[Ne~II] ratio between the 12\micron~sample and the BAT sample returns a $\sim 1.9\%$ probability of the null hypothesis, meaning that these samples are statistically different in their stellar content, with galaxies in the 12\micron~sample having  slightly smaller ratios on average, 1.1$\pm$0.9 , than that found for the BAT sample,  1.3$\pm$0.8. On the other hand, the K-S test for the [O~IV]/[Ne~III] ratio between the  the 12\micron~sample and the BAT sample returns a $\sim 26.5\%$ probability of the null hypothesis. These results suggest that even the 12\micron~sample could be subject to recent star formation contamination, as mapped through the [Ne~II] enhancement, in agreement with previous considerations \citep{1999ApJ...516..660H}. This mild  contamination due to star formation becomes noticeable when comparing 
the 12\micron~sample with the 14--195~keV sample, the latter of which is less  biased towards star-forming systems. Nonetheless, the similarity on their [O~IV]/[Ne~III] ratios suggest that both samples have a strong AGN contribution to their observed narrow line emission.

The tight correlation between [O~IV] and [Ne~V], evident in Figure~\ref{fig6}, is seen
here as well. The [Ne~V]/[Ne~II] versus [O~IV]/[Ne~II] plot shows a number of sources, including several
of the newly BAT-detected AGN and the two LINERs in the sample, with relatively weak high-ionization lines,
which can result from a weak AGN \citep[e.g.,][]{2004ApJ...614..558N}, strong nuclear star formation \citep[e.g.,][]{2008ApJ...689...95M}, extinction towards the NLR region \citep[e.g.,][]{2006A&A...453..525N}, or  shielding of the NLR from the ionizing continuum \citep[e.g.,][]{2007ApJ...656..148A}. Comparing the [Ne~III]/[Ne~II] to the [Ne~V]14.32$\micron$/[Ne~II] we found the same overall trend, however, the slope of the former  flattens towards lower ratios of [O~IV]/[Ne~II] possible due to the contribution of star formation to the [Ne~III].  Interestingly, all these ratios are log normal distributed with a standard deviation of only $\sim 0.3$~dex, except for the ${\rm [Ne~V]14.32\micron/[O~IV]}$ with a standard deviation of $\sim 0.2$~dex.  The [O~IV]/[Ne~II] emission line ratio was also compared with the redshift ($z$) in order to check if our previous results are been affected by aperture affects, in other words, if our results are biased toward small [Ne~III]/[Ne~II] and [O~IV]/[Ne~II] ratios at higher redshifts. We found that the Spearman rank, $\rho_s=0.129$; $P_\rho=0.30$,   and  Kendall test, $\tau=0.08$;$P_\tau=0.27$,  did not show any correlation  with $z$. A similar result was obtained for the [O~IV]/[Ne~III] ratios and the redshift where we found no correlation, with a Spearman rank, $\rho_s=-0.05$; $P_\rho=0.66$, and  Kendall test, $\tau=-0.05$;$P_\tau=0.51$.

\subsection{Extinction in the Mid-Infrared}
 
\begin{figure}
\epsscale{0.9}
\plotone{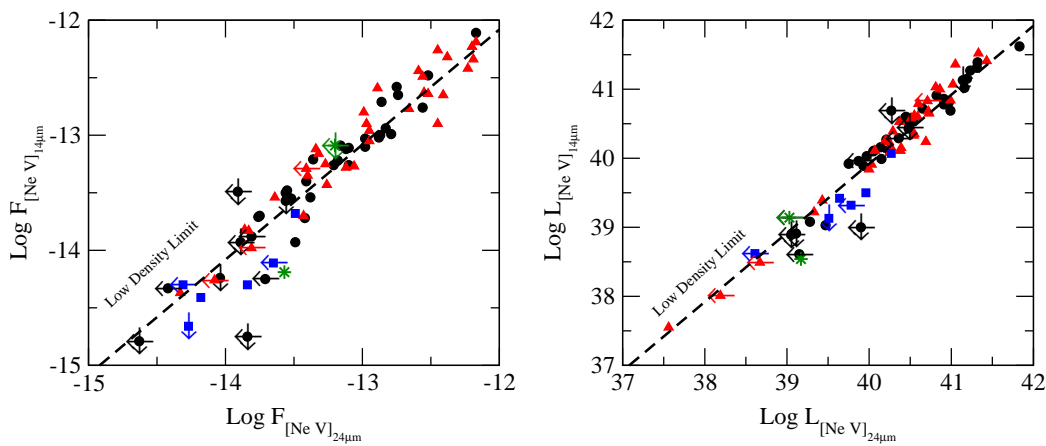}
\caption{Correlation between the ${\rm [Ne V]24.32\mu m}$ and ${\rm [Ne V]14.32\mu m}$ fluxes 
and luminosities for our X-ray selected sample. Symbols are the same as in Figure~\ref{fig4}. The low-density limit for this ratio is given by the dashed line; AGN lying
below and to the right have ${\rm [Ne V]14.32\mu m}/{\rm [Ne V]24.32\mu
m}$ ratios smaller than the theoretical limit.
The majority of the newly-detected BAT AGN have 
ratios below the limit, implying dust extinction towards 
the ${\rm [Ne V]14\mu m}$ emitting region.\label{fig8}}
\end{figure}

If wavelength-dependent extinction is important in the mid-infrared, it should be evident when comparing the
strengths of [Ne~V]~14.32$\mu$m and [Ne~V]~24.32$\mu$m. \cite{2007ApJ...664...71D} calculated the theoretical
values of this ratio as a function of density and electron temperature (see their Figure~1). As density
increases, the 24.32$\mu$m line is suppressed relative to the 14.32$\mu$m line, hence the most likely 
way in which the ratio can fall below the theoretical low density limit is extinction, suggesting that it  should
be stronger for the shorter-wavelength line. In Figure~\ref{fig8},
we show the correlation between the two [Ne~V] lines for the BAT sample, with the theoretical low density limit
over-plotted. We note that the correlation between the [Ne~V] emission lines is strong and is similar in both flux and luminosities, meaning that there is a true linear correlation (see Table~4).

\begin{figure}
\epsscale{0.9}
\plotone{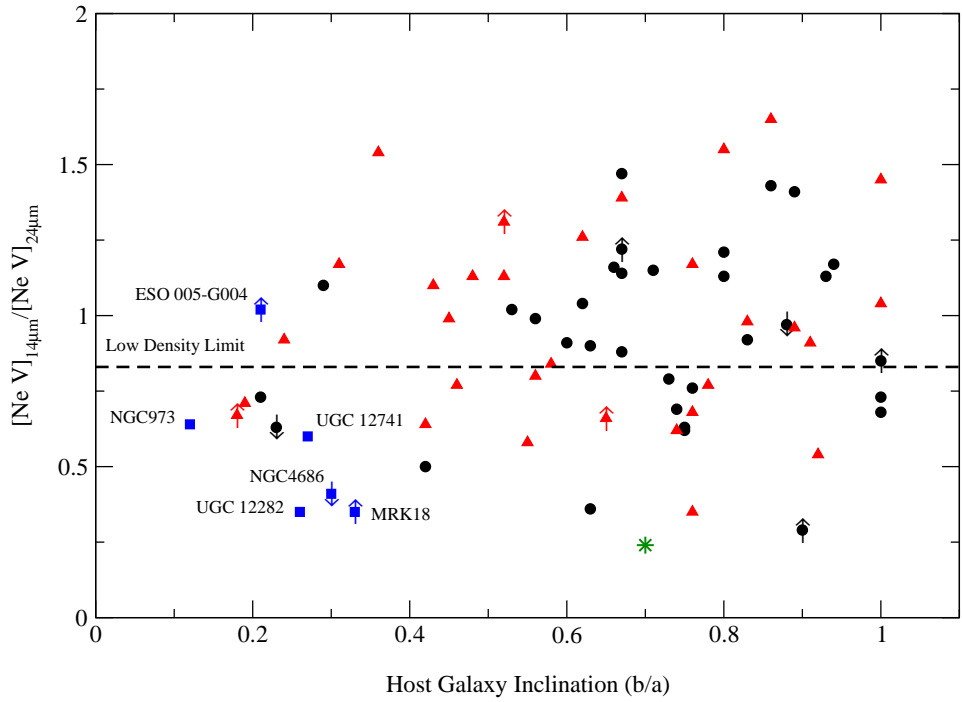}
\caption{The ${\rm [Ne V]_{24\mu m}}$ and ${\rm [Ne V]_{14\mu m}}$ ratios as
a function of host galaxy inclination for the BAT sample. Symbols are the same as in Figure~\ref{fig4}. The horizontal
line shows the low density limit. Note that, five out of six  of the newly detected BAT AGN 
reside in inclined hosts $b/a < 0.4$, hence host galaxy extinction may be responsible for their lack of AGN signatures in the optical\label{fig9}}
\end{figure}

For the BAT sample the average [Ne~V]$_{14.32 \micron}$/[Ne~V]$_{24.32\micron}$ ratio  is above the low density limit and similar for Seyfert 1 and Seyfert 2 galaxies, $1.0\pm0.3$   and $1.0\pm0.4$, respectively. The K-S test for this ratio returns a $\sim 94.0\%$ probability of the null hypothesis, i.e., that there is no difference between Seyfert 1 and Seyfert 2 galaxies, however, Seyfert 2 galaxies show a bigger dispersion. For a gas temperature of $10^4$~K, these results suggest an electron density of $n_e\approx 10^3$~cm$^{-3}$, in agreement with previous studies \citep[e.g.,][]{2002A&A...393..821S,2007ApJ...664...71D,2008ApJ...676..836T}.  On the other hand, five out of six of the newly BAT-detected AGN have ratios below the low density limit, with an average of $0.5\pm0.2$,  indicating possible dust extinction. In Figure~\ref{fig9}, we compared the [Ne~V] line ratio to the host galaxy inclination\footnote{The values for the major and minor diameter of the host galaxy, a and b respectively, were taken from NED}. All of the newly BAT-detected AGN are found in  inclined host galaxies which argues in favor of an scenario where  
the optical signatures of the AGN may have  been missed in this group, in part,  due to extinction in the plane of the host galaxy; as suggested by \cite{1990AJ.....99.1722K} and \cite{1997ApJ...489..615S} and more recently by GA09. However, one must note from Figure~\ref{fig9}, that there is no correlation between the  [Ne~V] line ratio and the value of the host galaxy inclination. Hence, host galaxy inclination cannot be solely  responsible for the observed  mid-infrared extinction in our sample.

In addition to providing some insight into the nature of the newly BAT-detected AGN, these plots show that $\sim 34\%$ of Seyfert 1 galaxies and $\sim 37\%$ of Seyfert 2 galaxies have [Ne~V] ratios below the low density limit, in good agreement with the percentage of Seyfert galaxies that fall  below the  low density limit for the neon ratio from the  high-resolution spectroscopy of the 12{$\mu$}m sample \citep{2010ApJ...709.1257T}.  If extinction at shorter wavelengths were responsible for the neon ratio below the low density limit,   it is plausible that the tightness of the [Ne~III] -- ${\rm L_{BAT}}$ correlation may be due to the combined effects of mid-infrared extinction and X-ray decrement due to Compton scatter in the BAT band \cite[e.g.,][]{2008ApJ...682...94M}. Similarly, the larger scatter for [O~IV] -- ${\rm L_{BAT}}$ may be  due, in part, to the lower extinction for [O~IV], compared to [Ne~III], hence the variations in the observed X-ray,  due to different  column densities toward the nuclear region,  become the dominant effect. One must note that the mid-infrared extinction curved by \cite{2001ApJ...548..296W} and \cite{2001ApJ...554..778L}  \citep[e.g, see Figure~16 in][]{2001ApJ...554..778L}, fails to  predict the required dust extinction at ${\rm A_{14.32\micron}/N_H}$ and ${\rm A_{24.32\micron}/N_H}$ to explain the observed [Ne~V] ratios  below the low density limit. Other possibilities to explain the observed neon ratios below the low density limit must be consider, such as, aperture effects and/or  adopted atomic data for neon, however, these possibilities  have also failed to fully explain the observed ratios below the low density limit \citep[e.g.,][]{2007ApJ...664...71D,2010ApJ...709.1257T}.

\section{Comparison between BAT AGN and other Sources}

Following the mid-infrared properties for the BAT 
AGN sample we investigated the relationship between different types of galaxies. 
For comparison we searched the literature for the most recent {\it Spitzer} high-resolution IRS spectroscopy, from this  we used the 
volume-limited sample of bolometrically luminous galaxies ($L_{IR} > 3 \times 10^9 {\rm L_\odot}$) to a distance of $D < 15~$Mpc of GA09 which includes H~II galaxies, LINERS, optically classified AGN 
and optically unclassified galaxies. We also used the recent atlas of starburst galaxies by B09 
that includes starburst galaxies with and without  evidence of an AGN. Finally,  we used a sample of 
extreme starburst galaxies which includes blue compact dwarfs (BCD) from H09. When a source 
was present in more than one compilation we gave priority to the data analysis and extraction presented in GA09, followed by B09 and H09.

Using [Ne~V]~14.32$\mu$m, GA09 identified AGN activity in
sources that were previously not classified by their optical spectra as AGN. While theoretical models predict 
that some fraction of [Ne~V] can be produce by energetic starbursts \citep[e.g.,][]{2008ApJ...678..686A}, H09 did not detect [Ne~V] emission   in their sample of extreme starburst galaxies. However, some  problems arise when looking at the [Ne~V] emission in  AGN. As we mentioned before,   
[Ne V]~14.32$\mu$m and 24.32$\mu$m  were not detected in all  the sources in our sample of BAT AGN. 
This suggests that because of dust extinction and instrumental sensitivity, 
[Ne~V] emission could go undetected even in intrinsically luminous  AGN. On the other hand, [Ne~II], [Ne~III] and [O~IV] are present 
in AGN, BCD, starburst, H~II galaxies and have a  wide range of ionization potentials and critical densities which allow us to 
study the connection between the active nucleus and star 
formation. One must note that H09 found that the most likely contributor to the [O~IV] emission in starburst and BCD galaxies is Wolf-Rayet 
stars. In this regard, and despite the fact that [O~IV] emission traces the AGN intrinsic luminosity in {\it pre-selected} AGN \citep{2008ApJ...682...94M}, [O~IV] cannot be solely associated with the power of the AGN.

\begin{figure}
\epsscale{0.9}
\plotone{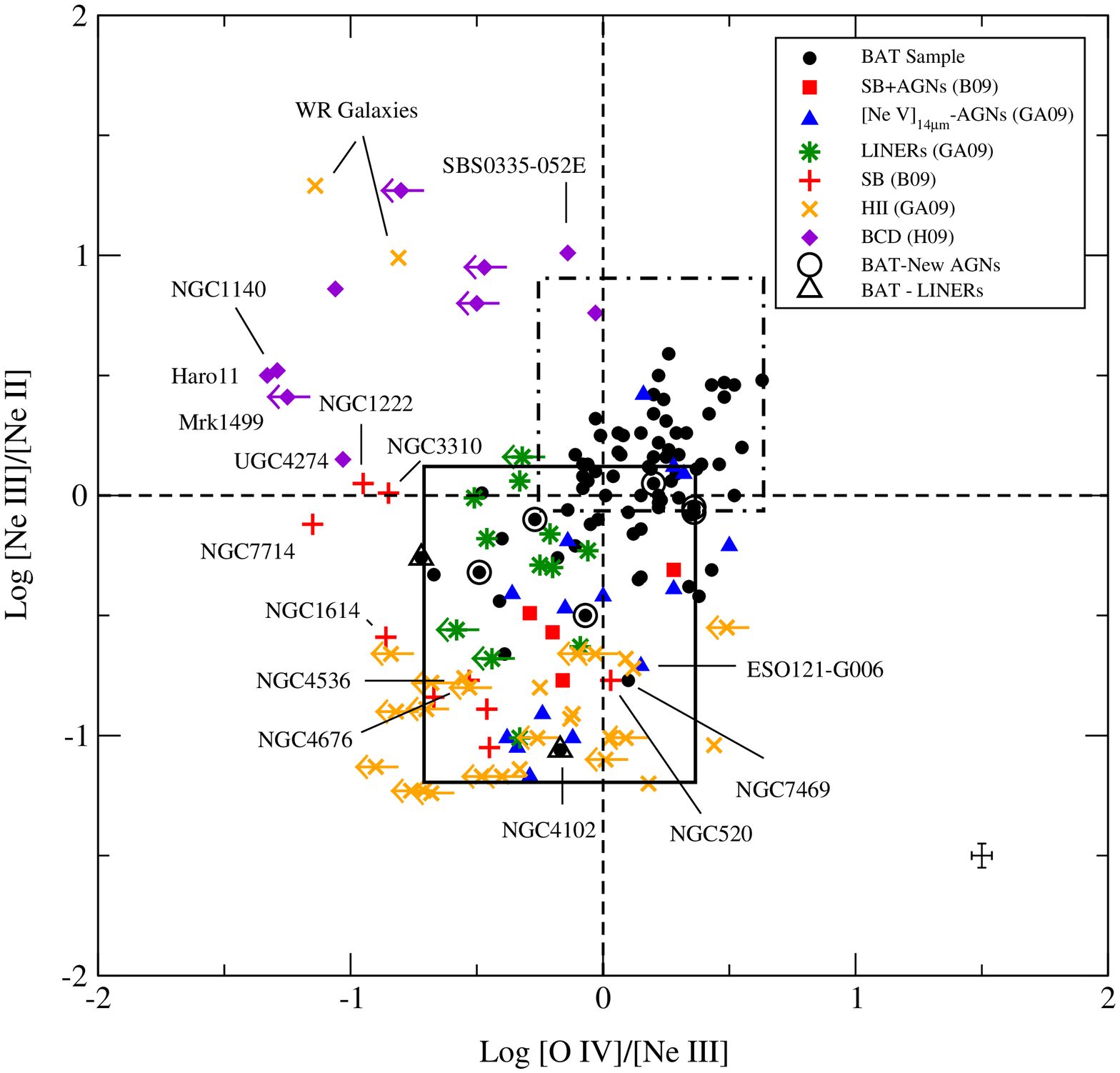}
\caption{[Ne~III]/[Ne~II] versus [O~IV]/[Ne~III] for a sample of different sources, including our BAT AGN, LINERs, starburst galaxies, HII galaxies and blue compact dwarf  from high-resolution {\it Spitzer} spectra from \cite{2009MNRAS.398.1165G} (GA09), \cite{2009ApJS..184..230B} (B09), and \cite{2009ApJ...704.1159H} (H09). Note that symbols are different than in  previous figures, see legend. The dashed lines represent a value of unity for the [Ne~III]/[Ne~II] and [O~IV]/[Ne~III] ratios. The rectangles represent the boundaries for the sample of  74 ultraluminous infrared galaxies ({\it solid line}) and 34 Palomar-Green  quasars ({\it point-dashed line}) from \cite{2009ApJS..182..628V}. 
 Interestingly, half of the BAT sample can be uniquely distinguished  from SB/HII/BCD galaxies by having  both [Ne~III]/[Ne~II] and [O~IV]/[Ne~III] ratios greater than unity. For the ratio errors, we propagated the uncertainty for individual emission lines  in quadrature. The error bars at the bottom right corner show an average error for the ratios presented in this plot. \label{fig11}}
\end{figure}

In Figure~\ref{fig11} we compare the [Ne~III]/[Ne~II] versus [O~IV]/[Ne~III] ratios. From this comparison 
there are four  distinctive regions to note: 1) the BAT AGN branch; 2) below the BAT AGN are the starburst (SB) and HII galaxies which have an apparent
[Ne~II] excess; 3) above and to the left of the BAT AGN are the BCD and two WR galaxies from GA09 (II~Zw40 and NGC~1569), and 4) LINERs, which seem to follow a connecting path between AGN and SB/HII galaxies but are mainly found in the lower-luminosity region of the BAT AGN branch. Despite the fact that [O~IV] has been detected in BCD, SB and HII galaxies, WR and BCD galaxies cannot produce [O~IV] as effectively as [Ne~III], unlike in AGN, making these sources have relative low [O~IV]/[Ne~III] ratios. On the other hand, the hardness of the photoionizing continuum from WR stars and other energetics phenomena, such as ULXs, can ionize ${\rm Ne^+}$ into ${\rm Ne^{2+}}$ resulting in relatively weak [Ne~II] \citep[e.g.,][]{2010ApJ...708..354B}, thus the  high [Ne~III]/[Ne~II] ratios observed in some BCD/SB galaxies.

There is also a small ``merge" region, at the lower ionization end of the BAT AGN branch,  that comprises LINERs and starburst galaxies with evidence for an AGN. This region  exemplify the different properties of LINERs, e.g., IR-luminous and -faint LINERs, with the latter having similar SED than that found in starburst galaxies \citep{2006ApJ...653L..13S}. Moreover, in this region we found five of the [Ne~V] AGN from GA09, 
some of which don't have optical signatures for an AGN.  Among these [Ne~V] AGN, NGC~660 and NGC~3628 have shown some evidence to harbor a LINER \citep{2004A&A...418..429F,2005ApJ...620..113D,2006ApJ...647..140F}. Also, based on this diagnostic, we found that   NGC~520, NGC~1614, NGC~4536, NGC~4676 could harbor an AGN, although these starburst galaxies don't show
optical evidence of such (B09;GA09). Of these, there is X-ray evidence for a Compton thick AGN in NGC~1614 \citep{2000A&A...357...13R}, [Ne~V]~14.31\micron~emission detected in NGC~4536 \citep{2008ApJ...677..926S} and NGC~4676 appears in the multi-wavelength LINER catalogue compiled by \cite{1999RMxAA..35..187C}. Interestingly,  these possible AGN are found in interacting systems or mergers with strong star formation. Perhaps, interacting or merger systems may provide enough fuel to the nuclear regions of the galaxy to trigger both the nuclear star formation and the AGN.

We also over plot in Figure~\ref{fig11} the sample of  74 ultraluminous infrared galaxies (ULIRGs) and 34 Palomar-Green  quasars (PG QSOs), observed with the IRS {\it Spitzer}, from \cite{2009ApJS..182..628V}. From this comparison its clear that ULIRGs show a wide range of  AGN and SB contribution to their mid-infrared emission lines, filling  the gap between the  BAT AGN and the SB/HII branch and  overlapping the small ``merge" region defined above, in agreement with the idea that ULIRGs are composite systems mainly powered by stellar activity \citep{2007ApJ...656..148A,2009ApJS..182..628V}. On the other hand, PG~QSOs overlap with the BAT AGN branch, in agreement with the high, typically larger than $\sim 80\%$, AGN contribution to their  bolometric luminosities \citep{2009ApJS..182..628V}. One must note that there are no  ULIRGs detected in the {\it Swift}-BAT catalog \citep{2010ApJS..186..378T,2010A&A...510A..48C}, except for the composite system NGC~6240 \citep{2006ApJ...640..204A}.

Moreover,  most of the non-BAT AGNs presented in this work, e.g., the SB+AGN sources from B09 and [Ne~V]~14.32$\mu$m-AGN  (GA09), in other words, AGN that have been selected because their optical and mid-infrared emission line properties, have smaller mid-infrared ratios than that found for the BAT AGN. In fact, half of the BAT sample can be uniquely distinguished  from SB/HII/BCD galaxies by having  both [Ne~III]/[Ne~II] and [O~IV]/[Ne~III] ratios greater than unity. This result suggests that the BAT sample represents a unique opportunity to study  high ionization  AGN, i.e., sources in which their optical/mid-infrared  emission signatures are dominated by the AGN. Figure~\ref{fig11} shows that  the [Ne~III]/[Ne~II] ratio as a stand alone  diagnostic may overestimate the fraction of AGN because Wolf-Rayet stars or another energetic phenomena  could produce [Ne~III]/[Ne~II] ratios even higher than those found on the BAT AGN. However,  by further constraining this ratio by using  [O~IV] we were able to separate AGN from energetic starbursts and other sources.

 As we mentioned before, caution must be taken when comparing fluxes between {\it Spitzer}  high-resolution orders because of the differences in aperture size. However, given the similarities between 
the ionization potentials for [Ne~III] and [O~IV], it is likely that both lines originate in similar regions, thus introducing less uncertainty in the use of the [O~IV]/[Ne~III] ratio. Nevertheless,  in HII/SB/BCD galaxies we could be underestimating the amount of [Ne~III] that is associated  with the same physical conditions that produce [O~IV], e.g., WR stars. This  will cause this ratio, when derived from equal apertures extractions,  to be displaced toward even smaller values  thus, enhancing the differences between these sources and our sample of hard X-ray selected AGN.

\section{Conclusions}

Using high-resolution {\it Spitzer}  IRS spectra, we have 
examined the mid-infrared emission-line properties of a sample of
hard X-ray selected AGN, detected by {\it Swift}/BAT. Our principle
conclusions are as follows.

1. The luminosity distribution for the mid-infrared emission lines and BAT continuum luminosities 
show no differences between Seyfert~1 and Seyfert~2 populations for the 
BAT sample.  The correlations between all the mid-infrared emission lines and BAT in both flux--flux and luminosity--luminosity are statistically significant, even when factoring the 
distance effect in luminosity-luminosity correlations.  The dispersion/tightness  
in these correlations is due to differences in the X-ray absorbing column 
densities, dust extinction and/or nuclear star formation activity. Moreover, the tight correlation found in the [Ne~III]-BAT relationship suggests that, on average,  there is no strong enhancement due to  star formation in the [Ne~III] emission in the BAT sample. 
Also, the slopes for the [Ne~III],[Ne~V] and [O~IV] versus BAT luminosities   
relationships are smaller in Seyfert~1 galaxies than in Seyfert~2s (which 
are around unity), which suggests that, while the amount of extinction towards the NLR is similar in both types,
the X-ray absorbing columns are large enough in Seyfert~2s to affect the hard X-ray
band, confirming the results of \cite{2008ApJ...682...94M} and \cite{2009ApJ...700.1878R}. This result  is in agreement with the  fact that the BAT/[O~IV] ratio statistically separates Seyfert~1s and Seyfert~2s.

2.  Although all of the correlations among the mid-infrared emission lines are 
strong, the worst correlations are for [Ne~V]-[Ne~II] 
and [O~IV]-[Ne~II], because of enhancement of the [Ne~II] from nuclear stellar activity \citep[see also][]{2008ApJ...689...95M}. 
While the tightness of 
these mid-infrared correlations suggests that dust extinction is not 
the driving physical process behind the mid-infrared relationships,
approximately $\sim$40$\%$ (including upper limits) of the sample have values for the ratio 
of the [Ne~V] emission lines below the low-density theoretical limit, suggesting dust extinction as the physical process 
responsible. Exploring this, we found that all of the newly discovered BAT AGNs in our sample, which are under-luminous in [O~IV] and [Ne~V]14/24$\mu$m, are 
 found on  inclined host galaxies, and all but one have [Ne~V] 
ratios below the critical density limit. Hence, it is likely that the newly found BAT  AGN in our sample 
lack optical AGN signatures because of host galaxy extinction towards their 
NLRs. However the lack of correlation between host galaxy inclination and the neon ratios suggest that extinction along the plane of the host galaxy cannot be responsible for  the observed extinction in all the BAT AGN sample.

3. We compared the BAT AGNs with different starburst and H~II galaxies,
so-called [Ne~V] active galaxies, and LINERs \citep{2009MNRAS.398.1165G,2009ApJS..184..230B,2009ApJ...704.1159H}. We found that the BAT AGN   fall into a distinctive region based
on the [Ne~III]/[Ne~II] and [O~IV]/[Ne~III] ratios. Using 
[Ne~III] and [O~IV] emission, previously connected with  AGN power \citep[e.g.,][]{2007ApJ...655L..73G,2008ApJ...682...94M}, 
does not unambiguously identify AGNs as an stand alone diagnostic 
because Wolf-Rayet stars or another energetic phenomena (perhaps ULXs) could enhance the 
observed emission. While it is likely that detection of [Ne~V] indicates the presence of an AGN, the strongest of the [Ne~V] lines have $\sim$1/3  less flux than [O~IV] an thus will be more difficult to detect in weak or faint AGN. Therefore, our composite method  using the  [Ne~II], [Ne~III] 
and [O~IV], represents a strong  and simple diagnostic by using only three emission lines to identify an AGN. Based on this, we found that   NGC~520, NGC~1614, NGC~4536, NGC~4676 could harbor an AGN, although these starburst galaxies don't show
optical evidence of such (B09;GA09). Of these, there is X-ray evidence for a Compton thick AGN in NGC~1614 \citep{2000A&A...357...13R}, [Ne~V]~14.31\micron~emission detected in NGC~4536 \citep{2008ApJ...677..926S} and NGC~4676 appears in the multi-wavelength LINER catalogue compiled by \cite{1999RMxAA..35..187C}.  Such line  diagnostic will be particularly useful to analyze spectra from new IR missions, such as the {\it James Webb Space Telescope} \citep{2006SSRv..123..485G}.

We also found that ULIRGs and PG QSOs occupy two distinctive regions in our emission line diagnostic.  Most ULIRGs fall into  the gap between the  BAT AGN and the SB/HII branch, in agreement with the idea that ULIRGs are composite systems mainly powered by stellar activity \citep{2007ApJ...656..148A,2009ApJS..182..628V}. On the other hand,  PG QSOs overlap with the BAT AGN branch, in agreement with the high, typically larger than $\sim 80\%$, AGN contribution to their  bolometric luminosities \citep{2009ApJS..182..628V}.

Finally, most of the non-BAT AGNs presented in our study, AGN that have been selected because their optical and mid-infrared emission line properties, have smaller mid-infrared ratios than that found for the BAT AGN. In this regard, half of the BAT sample can be uniquely distinguished  from SB/HII/BCD galaxies by having  both [Ne~III]/[Ne~II] and [O~IV]/[Ne~III] ratios greater than unity. Moreover, when comparing  the  12\micron~and our BAT selected AGN we found that the  [Ne~III]/[Ne~II] ratio distribution between the samples  is statistically different with   sources in the 12\micron~sample having on average   lower ratios than that found in the BAT AGN, or alternatively higher recent stellar activity. This mild  contamination due to star formation becomes noticeable when comparing 
the 12\micron~sample with the 14--195~keV sample, the latter of which is less  biased towards star-forming systems.    Despite the fact that  both samples have a strong AGN contribution to their observed narrow line emission, this result suggests that the BAT sample represents a unique opportunity to study high ionization  AGN, sources in which their optical/mid-infrared  emission signatures are dominated by the AGN, thus, providing  the most representative  sample in terms of galaxy population  and stellar content.


\acknowledgments
We thank the referee  for very useful comments that improved
the paper. We also would also like to acknowledge  S. Veilleux for valuable comments on the work. This research was supported by an appointment to the NASA Postdoctoral Program at the Goddard Space Flight Center, administered by Oak Ridge Associated Universities through a contract with NASA. L.M.W. acknowledges support under Hubble Fellowship program number HST-HF-51263.01-A from the Space Telescope Science Institute, which is 
operated by the Association of Universities for Research in Astronomy, Incorporated, under NASA contract NAS5-26555. This research has made use of the NASA/IPAC Extragalactic Database (NED) which is operated by the Jet Propulsion Laboratory, 
California Institute of Technology, under contract with the National Aeronautics and Space Administration. The IRS was a collaborative venture between 
Cornell University and Ball Aerospace Corporation funded by NASA through the Jet Propulsion Laboratory and Ames Research Center. SMART was developed by the IRS Team at Cornell University and is available through the {\it Spitzer} Science Center at Caltech.




\clearpage
\bibliographystyle{apj}
\bibliography{ms} 


\clearpage

\clearpage



\clearpage

\clearpage

\clearpage

\clearpage

\clearpage

\clearpage

\clearpage

\clearpage

\begin{deluxetable}{lcccccccc}
\rotate
\tabletypesize{\scriptsize}
\tablewidth{0pt}
\tablecaption{{\it Spitzer} IRS High-Resolution Spectroscopy of the BAT sample of AGNs\label{table1}}

\tablehead{ 
\colhead{Name} & \colhead{Distance}&\colhead{Type\tablenotemark{a}}&\colhead{BAT\tablenotemark{b}}&\colhead{[Ne~II]}&\colhead{[Ne V]} &\colhead{[Ne~III]}& \colhead{[Ne V]} &\colhead{[O~IV]} \\
  & (Mpc)  &  &  &(12.81$\mu$m)& (14.32$\mu$m)& (15.56$\mu$m)& (24.32$\mu$m)& (25.89$\mu$m)\\
  \cline{5-9}\\
  &  &  &  & \multicolumn{5}{c}{Integrated Line Fluxes ($10^{-21}{\rm W~cm^{-2}}$)}\\
}

\startdata

2MASX J05580206-3820043	&	146.8	&	1	&	3.99	$\pm$	0.37	&	3.20	$\pm$	0.89	&	1.89	$\pm$	0.13	&	4.72	$\pm$	0.14	&	3.80	$\pm$	0.51	&	3.69	$\pm$	0.45	\\
3C120	&	143.0	&	1	&	11.89	$\pm$	0.63	&	9.19	$\pm$	0.66	&	17.21	$\pm$	0.28	&	27.43	$\pm$	0.92	&	27.63	$\pm$	8.92	&	116.68	$\pm$	1.22	\\
Ark 120	&	141.7	&	1	&	7.08	$\pm$	0.57	&	3.47	$\pm$	0.91	&	$<$1.18			&	4.33	$\pm$	0.79	&	$<$1.29			&	4.03	$\pm$	0.38	\\
Cen A\tablenotemark{c,g}	&	4.9	&	2	&	92.62	$\pm$	0.71	&	193.00			&	23.16			&	147.65			&	29.92			&	131.24			\\
ESO 005-G004\tablenotemark{g}	&	22.4	&	\nodata	&	4.48	$\pm$	0.60	&	16.64	$\pm$	1.19	&	0.50	$\pm$	0.12	&	5.29	$\pm$	0.34	&	$<$0.49			&	4.49	$\pm$	0.19	\\
ESO 033- G002	&	77.5	&	2	&	2.61	$\pm$	0.45	&	2.67	$\pm$	0.18	&	5.57	$\pm$	0.56	&	8.34	$\pm$	0.56	&	5.37	$\pm$	0.51	&	13.85	$\pm$	0.73	\\
ESO 103-G035	&	56.7	&	2	&	11.14	$\pm$	0.59	&	30.92	$\pm$	2.07	&	15.85	$\pm$	3.49	&	41.62	$\pm$	1.78	&	10.31	$\pm$	0.87	&	34.40	$\pm$	0.75	\\
ESO 140-G043 	&	60.5	&	1	&	4.57	$\pm$	0.66	&	11.41	$\pm$	0.26	&	7.83	$\pm$	0.70	&	13.93	$\pm$	0.32	&	7.95	$\pm$	0.44	&	27.46	$\pm$	0.35	\\
ESO 323-G077	&	64.2	&	1.2	&	4.70	$\pm$	0.66	&	40.16	$\pm$	1.41	&	5.44	$\pm$	1.61	&	18.07	$\pm$	1.07	&	6.19	$\pm$	0.14	&	25.11	$\pm$	0.62	\\
ESO 362-G018	&	53.1	&	1.5	&	6.22	$\pm$	0.52	&	9.94	$\pm$	0.23	&	3.15	$\pm$	0.37	&	7.16	$\pm$	0.18	&	2.75	$\pm$	0.20	&	10.11	$\pm$	0.37	\\
ESO 417- G006	&	69.7	&	2	&	3.06	$\pm$	0.46	&	1.77	$\pm$	0.14	&	0.42	$\pm$	0.10	&	2.54	$\pm$	0.07	&	0.47	$\pm$	0.16	&	4.04	$\pm$	0.18	\\
F9	&	205.9	&	1	&	5.07	$\pm$	0.45	&	2.91	$\pm$	0.68	&	$<$2.66			&	5.31	$\pm$	0.37	&	2.74	$\pm$	0.37	&	6.08	$\pm$	0.26	\\
F49	&	86.2	&	2	&	2.93	$\pm$	0.54	&	39.55	$\pm$	2.93	&	25.50	$\pm$	2.96	&	47.12	$\pm$	2.43	&	12.74	$\pm$	0.36	&	39.22	$\pm$	2.37	\\
IC 486	&	114.8	&	1	&	3.22	$\pm$	0.70	&	6.86	$\pm$	0.05	&	3.30	$\pm$	0.20	&	6.79	$\pm$	0.21	&	2.84	$\pm$	0.01	&	11.19	$\pm$	0.29	\\
IC 1816	&	72.5	&	1	&	2.58	$\pm$	0.48	&	18.36	$\pm$	1.51	&	6.21	$\pm$	0.61	&	19.64	$\pm$	1.17	&	4.34	$\pm$	0.30	&	16.52	$\pm$	1.81	\\
IC 4329A	&	68.7	&	1.2	&	33.08	$\pm$	0.62	&	30.19	$\pm$	2.76	&	32.79	$\pm$	6.01	&	65.30	$\pm$	0.95	&	29.90	$\pm$	1.74	&	103.58	$\pm$	3.24	\\
IC 5063	&	48.3	&	2	&	8.59	$\pm$	0.72	&	28.22	$\pm$	3.34	&	35.97	$\pm$	1.58	&	73.67	$\pm$	4.61	&	25.86	$\pm$	1.20	&	117.21	$\pm$	11.13	\\
MCG-01-13-025\tablenotemark{f}	&	68.0	&	1.2	&	$<$4.5			&	2.87	$\pm$	0.18	&	$<$0.18			&	1.90	$\pm$	0.05	&	$<$1.44			&	0.76	$\pm$	0.18	\\
MCG-05-23-016	&	36.1	&	2	&	20.77	$\pm$	0.56	&	18.13	$\pm$	0.42	&	11.06	$\pm$	0.50	&	16.96	$\pm$	0.93	&	11.19	$\pm$	0.26	&	27.95	$\pm$	7.61	\\
MCG-01-24-012	&	84.2	&	2	&	4.58	$\pm$	0.51	&	6.74	$\pm$	0.41	&	2.90	$\pm$	0.27	&	6.07	$\pm$	0.36	&	2.30	$\pm$	0.58	&	10.07	$\pm$	0.77	\\
MCG-02-58-22	&	205.1	&	1.5	&	10.17	$\pm$	0.57	&	5.29	$\pm$	0.39	&	2.80	$\pm$	0.34	&	9.71	$\pm$	0.16	&	3.05	$\pm$	0.39	&	13.74	$\pm$	1.74	\\
MCG-03-34-064	&	70.8	&	1.8	&	3.15	$\pm$	0.45	&	57.83	$\pm$	7.29	&	54.98	$\pm$	5.26	&	120.13	$\pm$	6.88	&	35.46	$\pm$	1.92	&	110.83	$\pm$	9.35	\\
MCG-06-30-015	&	32.9	&	1.2	&	7.82	$\pm$	0.57	&	4.20	$\pm$	0.12	&	6.05	$\pm$	0.76	&	6.62	$\pm$	0.33	&	6.62	$\pm$	0.05	&	23.49	$\pm$	0.81	\\
MRK 3\tablenotemark{c}	&	57.7	&	2	&	15.65	$\pm$	0.61	&	100.00			&	64.50			&	179.00			&	67.50			&	214.00			\\
MRK 6\tablenotemark{d}	&	80.6	&	1.5	&	7.61	$\pm$	0.55	&	28.00	$\pm$	0.23	&	9.39	$\pm$	0.19	&	49.34	$\pm$	0.32	&	10.43	$\pm$	0.21	&	48.24	$\pm$	0.27	\\
MRK 18\tablenotemark{f}	&	47.2	&	\nodata	&	$<$3.1			&	17.04	$\pm$	0.71	&	0.78	$\pm$	0.01	&	8.24	$\pm$	0.24	&	$<$2.25			&	2.69	$\pm$	0.14	\\
MRK 79	&	95.3	&	1.2	&	4.89	$\pm$	0.52	&	11.29	$\pm$	3.71	&	9.62	$\pm$	1.07	&	20.43	$\pm$	0.50	&	13.16	$\pm$	1.10	&	39.99	$\pm$	2.68	\\
MRK 335	&	111.1	&	1.2	&	2.47	$\pm$	0.41	&	2.10	$\pm$	0.31	&	1.31	$\pm$	0.11	&	2.70	$\pm$	0.15	&	$<$1.55			&	6.31	$\pm$	0.17	\\
MRK 348	&	64.2	&	2	&	13.66	$\pm$	0.56	&	15.34	$\pm$	0.74	&	6.90	$\pm$	0.36	&	20.60	$\pm$	0.79	&	4.75	$\pm$	0.27	&	17.87	$\pm$	0.23	\\
MRK 352	&	63.5	&	1	&	4.16	$\pm$	0.50	&	0.54	$\pm$	0.08	&	$<$0.16			&	0.34	$\pm$	0.02	&	$<$0.23			&	0.26	$\pm$	0.04	\\
MRK 509	&	149.2	&	1.2	&	9.44	$\pm$	0.68	&	11.98	$\pm$	1.06	&	7.61	$\pm$	1.74	&	17.32	$\pm$	3.29	&	7.60	$\pm$	0.10	&	27.54	$\pm$	0.42	\\
MRK 590\tablenotemark{f}	&	112.7	&	1.2	&	$<$3.7			&	5.26	$\pm$	0.46	&	$<$3.22			&	5.42	$\pm$	0.12	&	$<$1.23			&	1.79	$\pm$	0.26	\\
MRK 766	&	55.2	&	1.5	&	2.42	$\pm$	0.29	&	24.27	$\pm$	0.94	&	22.14	$\pm$	0.23	&	23.52	$\pm$	1.41	&	18.32	$\pm$	1.60	&	46.47	$\pm$	0.84	\\
MRK 817	&	136.1	&	1.5	&	2.21	$\pm$	0.36	&	4.57	$\pm$	0.88	&	2.88	$\pm$	0.48	&	5.51	$\pm$	0.77	&	4.22	$\pm$	0.73	&	6.06	$\pm$	0.23	\\
MRK 841	&	158.2	&	1.5	&	2.93	$\pm$	0.37	&	3.18	$\pm$	0.18	&	8.12	$\pm$	0.31	&	12.46	$\pm$	0.73	&	6.96	$\pm$	0.33	&	22.91	$\pm$	1.08	\\
NGC 454\tablenotemark{f}	&	51.8	&	2	&	$<$2.3			&	4.70	$\pm$	0.40	&	3.73	$\pm$	0.14	&	6.40	$\pm$	0.17	&	5.54	$\pm$	0.31	&	15.80	$\pm$	0.85	\\
NGC 513	&	83.8	&	2	&	2.06	$\pm$	0.43	&	10.38	$\pm$	1.78	&	1.52	$\pm$	0.23	&	4.71	$\pm$	0.32	&	1.38	$\pm$	0.21	&	6.69	$\pm$	0.41	\\
NGC 788	&	58.1	&	2	&	9.33	$\pm$	0.57	&	6.59	$\pm$	0.11	&	5.35	$\pm$	0.34	&	13.60	$\pm$	0.50	&	8.68	$\pm$	1.28	&	24.14	$\pm$	0.36	\\
NGC 931	&	71.2	&	1.5	&	6.56	$\pm$	0.50	&	5.50	$\pm$	0.59	&	9.91	$\pm$	2.31	&	15.94	$\pm$	0.52	&	13.52	$\pm$	0.58	&	42.90	$\pm$	0.63	\\
NGC 973	&	69.3	&	\nodata	&	3.09	$\pm$	0.58	&	8.03	$\pm$	0.09	&	2.07	$\pm$	0.18	&	8.92	$\pm$	0.38	&	3.21	$\pm$	0.27	&	14.11	$\pm$	2.64	\\
NGC 1052\tablenotemark{g}	&	17.8	&	LINER	&	3.75	$\pm$	0.67	&	22.22	$\pm$	0.94	&	0.64	$\pm$	0.03	&	12.31	$\pm$	0.48	&	2.69	$\pm$	1.33	&	2.37	$\pm$	0.10	\\
NGC 1194	&	58.0	&	1.9	&	3.64	$\pm$	0.60	&	3.27	$\pm$	0.34	&	4.48	$\pm$	0.12	&	8.27	$\pm$	0.41	&	3.98	$\pm$	0.32	&	14.39	$\pm$	0.10	\\
NGC 1365\tablenotemark{g}	&	18.9	&	1.8	&	7.19	$\pm$	0.44	&	161.67	$\pm$	17.48	&	22.35	$\pm$	1.97	&	61.06	$\pm$	0.90	&	38.53	$\pm$	1.50	&	145.38	$\pm$	8.89	\\
NGC 2110	&	33.1	&	2	&	35.01	$\pm$	0.70	&	60.19	$\pm$	5.34	&	5.22	$\pm$	0.82	&	47.40	$\pm$	0.71	&	7.65	$\pm$	0.63	&	45.71	$\pm$	3.41	\\
NGC 2992\tablenotemark{g}	&	30.5	&	2	&	4.82	$\pm$	0.63	&	53.65	$\pm$	3.66	&	32.62	$\pm$	5.38	&	61.06	$\pm$	1.98	&	27.81	$\pm$	0.20	&	114.22	$\pm$	6.44	\\
NGC 3079\tablenotemark{c,g}	&	20.4	&	2	&	3.44	$\pm$	0.44	&	104.00			&	1.04			&	22.88			&	$<$1.56			&	9.26			\\
NGC 3081\tablenotemark{g}	&	32.5	&	2	&	10.24	$\pm$	0.67	&	12.62	$\pm$	1.16	&	12.62	$\pm$	0.81	&	36.46	$\pm$	1.25	&	35.79	$\pm$	0.05	&	119.73	$\pm$	8.42	\\
NGC 3227\tablenotemark{g}	&	20.6	&	1.5	&	14.13	$\pm$	0.50	&	65.05	$\pm$	6.91	&	26.23	$\pm$	1.23	&	74.62	$\pm$	2.00	&	17.82	$\pm$	0.99	&	64.91	$\pm$	2.31	\\
NGC 3281	&	45.5	&	2	&	9.01	$\pm$	0.66	&	19.94	$\pm$	2.16	&	47.51	$\pm$	2.75	&	58.35	$\pm$	2.48	&	42.15	$\pm$	3.24	&	174.65	$\pm$	13.25	\\
NGC 3516\tablenotemark{d,g}	&	38.9	&	1.5	&	12.54	$\pm$	0.45	&	8.07	$\pm$	0.25	&	7.88	$\pm$	0.50	&	17.72	$\pm$	0.33	&	10.39	$\pm$	0.33	&	46.92	$\pm$	0.35	\\
NGC 3783\tablenotemark{g}	&	38.5	&	1	&	19.45	$\pm$	0.66	&	19.82	$\pm$	0.79	&	19.54	$\pm$	3.44	&	26.18	$\pm$	0.57	&	13.82	$\pm$	0.29	&	39.25	$\pm$	0.07	\\
NGC 4051\tablenotemark{g}	&	17.1	&	1.5	&	4.34	$\pm$	0.35	&	19.69	$\pm$	0.92	&	10.17	$\pm$	0.65	&	16.35	$\pm$	0.44	&	16.19	$\pm$	2.53	&	36.95	$\pm$	2.05	\\
NGC 4102\tablenotemark{g}	&	17.0	&	LINER	&	2.58	$\pm$	0.38	&	349.81	$\pm$	39.88	&	$<$8.10			&	30.44	$\pm$	1.12	&	$<$6.29			&	20.66	$\pm$	7.69	\\
NGC 4138\tablenotemark{g}	&	17.0	&	1.9	&	3.69	$\pm$	0.45	&	3.23	$\pm$	0.52	&	0.54	$\pm$	0.10	&	2.82	$\pm$	0.37	&	$<$0.83			&	2.03	$\pm$	0.18	\\
NGC 4151\tablenotemark{c,g}	&	20.3	&	1.5	&	62.23	$\pm$	0.46	&	134.00			&	77.72			&	204.35			&	67.67			&	236.51			\\
NGC 4235\tablenotemark{g}	&	35.1	&	1	&	2.40	$\pm$	0.55	&	3.69	$\pm$	0.65	&	$<$0.58			&	3.70	$\pm$	0.54	&	0.92	$\pm$	0.19	&	3.77	$\pm$	0.77	\\
NGC 4388\tablenotemark{g}	&	16.8	&	2	&	34.64	$\pm$	0.52	&	79.74	$\pm$	4.76	&	45.35	$\pm$	0.84	&	108.18	$\pm$	1.56	&	64.10	$\pm$	0.12	&	311.42	$\pm$	25.79	\\
NGC 4395\tablenotemark{g}	&	3.9	&	1.8	&	3.12	$\pm$	0.41	&	4.74	$\pm$	0.25	&	1.47	$\pm$	0.13	&	7.02	$\pm$	0.58	&	1.49	$\pm$	0.62	&	8.16	$\pm$	0.26	\\
NGC 4507	&	50.3	&	2	&	22.51	$\pm$	0.68	&	33.73	$\pm$	2.63	&	12.50	$\pm$	1.39	&	28.63	$\pm$	2.36	&	10.72	$\pm$	1.96	&	36.33	$\pm$	4.14	\\
NGC 4593\tablenotemark{g}	&	39.5	&	1	&	9.79	$\pm$	0.62	&	8.31	$\pm$	0.30	&	5.56	$\pm$	1.27	&	7.89	$\pm$	0.59	&	8.02	$\pm$	1.96	&	13.39	$\pm$	2.00	\\
NGC 4686	&	71.6	&	\nodata	&	3.08	$\pm$	0.45	&	2.13	$\pm$	0.16	&	$<$0.22			&	1.71	$\pm$	0.13	&	0.53	$\pm$	0.09	&	0.91	$\pm$	0.30	\\
NGC 526A	&	81.9	&	1.5	&	5.96	$\pm$	0.51	&	6.22	$\pm$	1.61	&	3.94	$\pm$	0.47	&	9.59	$\pm$	1.10	&	3.87	$\pm$	0.86	&	17.34	$\pm$	0.84	\\
NGC 5506\tablenotemark{g}	&	28.7	&	1.9	&	25.64	$\pm$	0.50	&	91.75	$\pm$	3.31	&	58.28	$\pm$	3.34	&	152.13	$\pm$	9.13	&	63.25	$\pm$	2.40	&	252.82	$\pm$	2.29	\\
NGC 5548	&	73.5	&	1.5	&	8.08	$\pm$	0.50	&	8.93	$\pm$	0.70	&	1.95	$\pm$	0.10	&	8.99	$\pm$	0.97	&	1.73	$\pm$	0.75	&	12.75	$\pm$	0.80	\\
NGC 5728\tablenotemark{g}	&	42.2	&	2	&	10.54	$\pm$	0.71	&	30.44	$\pm$	1.81	&	23.56	$\pm$	0.77	&	54.76	$\pm$	0.51	&	28.15	$\pm$	0.56	&	118.40	$\pm$	7.61	\\
NGC 5995	&	108.5	&	2	&	4.51	$\pm$	0.61	&	13.32	$\pm$	1.91	&	7.60	$\pm$	1.42	&	9.19	$\pm$	0.59	&	4.61	$\pm$	0.84	&	12.20	$\pm$	0.80	\\
NGC 6240\tablenotemark{e}	&	105.4	&	2	&	7.30	$\pm$	0.62	&	193.10	$\pm$	3.70	&	5.10	$\pm$	0.90	&	70.40	$\pm$	2.40	&	$<$3.90			&	27.20	$\pm$	0.70	\\
NGC 6860	&	63.6	&	1	&	6.50	$\pm$	0.73	&	5.88	$\pm$	0.11	&	2.95	$\pm$	0.94	&	7.50	$\pm$	0.53	&	2.84	$\pm$	0.21	&	11.70	$\pm$	0.39	\\
NGC 7172\tablenotemark{g}	&	33.9	&	2	&	18.11	$\pm$	0.70	&	32.03	$\pm$	2.44	&	8.99	$\pm$	0.80	&	15.71	$\pm$	0.69	&	11.26	$\pm$	0.95	&	42.60	$\pm$	3.44	\\
NGC 7213\tablenotemark{g}	&	22.0	&	1.5	&	5.75	$\pm$	0.67	&	27.47	$\pm$	1.33	&	0.56	$\pm$	0.01	&	12.77	$\pm$	0.66	&	$<$1.94			&	2.75	$\pm$	0.59	\\
NGC 7314\tablenotemark{g}	&	18.3	&	1.9	&	4.63	$\pm$	0.59	&	8.97	$\pm$	0.74	&	16.84	$\pm$	0.60	&	23.28	$\pm$	0.41	&	21.92	$\pm$	0.40	&	69.62	$\pm$	7.82	\\
NGC 7469\tablenotemark{c}	&	69.8	&	1.2	&	6.66	$\pm$	0.44	&	200.00			&	11.60			&	34.00			&	14.70			&	43.00			\\
NGC 7582\tablenotemark{g}	&	22.0	&	2	&	7.92	$\pm$	0.55	&	250.94	$\pm$	3.53	&	38.02	$\pm$	3.86	&	104.99	$\pm$	4.37	&	59.54	$\pm$	6.39	&	227.65	$\pm$	10.88	\\
NGC 7603	&	127.6	&	1.5	&	4.70	$\pm$	0.51	&	9.32	$\pm$	0.72	&	0.47	$\pm$	0.01	&	5.07	$\pm$	0.49	&	$<$0.38			&	3.34	$\pm$	0.18	\\
NGC 7682	&	73.4	&	2	&	2.27			&	5.46	$\pm$	0.25	&	1.98	$\pm$	0.19	&	8.07	$\pm$	0.15	&	3.69	$\pm$	1.46	&	16.21	$\pm$	0.66	\\
UGC 03601	&	73.3	&	1.5	&	4.38	$\pm$	0.67	&	5.26	$\pm$	0.53	&	2.01	$\pm$	0.12	&	7.60	$\pm$	0.13	&	1.79	$\pm$	0.44	&	13.45	$\pm$	0.65	\\
UGC 06728	&	27.7	&	1.2	&	2.95	$\pm$	0.37	&	1.40	$\pm$	0.36	&	1.16	$\pm$	0.10	&	1.38	$\pm$	0.05	&	3.24	$\pm$	0.03	&	4.61	$\pm$	0.81	\\
UGC 12282	&	72.7	&	\nodata	&	2.49	$\pm$	0.50	&	2.08	$\pm$	0.28	&	0.50	$\pm$	0.01	&	1.76	$\pm$	0.05	&	1.44	$\pm$	0.20	&	4.07	$\pm$	0.37	\\
UGC 12741	&	74.7	&	\nodata	&	4.00	$\pm$	0.59	&	1.72	$\pm$	0.07	&	0.39	$\pm$	0.06	&	1.55	$\pm$	0.05	&	0.65	$\pm$	0.06	&	3.58	$\pm$	0.38	\\

\enddata
\tablenotetext{a}{AGN types are taken from NED.} 
\tablenotetext{b}{The BAT flux is presented in units of ${\rm 10^{-11}ergs~s^{-1}cm^{-2}}$.} 
\tablenotetext{c}{\cite{2005ApJ...633..706W}}
\tablenotetext{d}{\cite{2008ApJ...676..836T}}
\tablenotetext{e}{\cite{2006ApJ...640..204A}}
\tablenotetext{f}{From the 9-month BAT survey \citep{2008ApJ...681..113T}}
\tablenotetext{g}{Distances are taken from EDD}
\end{deluxetable}

\clearpage

\begin{deluxetable}{cccccccc}
\tabletypesize{\small}
\tablewidth{0pt} 
\tablecaption{Statistical Analysis for the Different  Relationships Between the Mid-infrared Emission Lines and BAT Luminosity for the  Sample}
\tablehead{ \colhead{Variables}   &\colhead{$\rho_s$}&\colhead{$P_\rho$}&\colhead{$\tau$}&\colhead{$P_\tau$}&\colhead{$\tau_p$}
&\colhead{$\sigma$} &\colhead{$P_\tau (p)$} }
\startdata 
{\bf All Sample}\\

BAT -- [Ne~II] & 0.66 & $<1\times 10^{-6}$ & 0.49 & $<1\times 10^{-6}$ & 0.37 & 0.07 & $<1\times 10^{-6}$ \\
BAT -- [Ne~III] & 0.76 & $<1\times 10^{-6}$ & 0.58 & $<1\times 10^{-6}$ & 0.45 & 0.06 & $<1\times 10^{-6}$ \\
BAT -- [O~IV] & 0.65 & $<1\times 10^{-6}$ & 0.48 & $<1\times 10^{-6}$ & 0.36 & 0.06 & $<1\times 10^{-6}$ \\
BAT -- [Ne~V]$_{14\micron}$ & 0.63 & $<1\times 10^{-6}$ & 0.50 & $<1\times 10^{-6}$ & 0.38 & 0.07 & $<1\times 10^{-6}$ \\
{\rm  [NeIII] -- [NeII]} & 0.89 & $<1\times 10^{-6}$ & 0.74 & $<1\times 10^{-6}$ & 0.68 & 0.06 & $<1\times 10^{-6}$ \\
{\rm [O~IV] -- [Ne~II]} & 0.80 & $<1\times 10^{-6}$ & 0.60 & $<1\times 10^{-6}$ & 0.54 & 0.06 & $<1\times 10^{-6}$ \\
{\rm [O~IV] -- [Ne~III]} & 0.94 & $<1\times 10^{-6}$ & 0.80 & $<1\times 10^{-6}$ & 0.76 & 0.06 & $<1\times 10^{-6}$ \\
{\rm [Ne~V]$_{14\micron}$  -- [Ne~II]}& 0.77 & $<1\times 10^{-6}$ & 0.59 & $<1\times 10^{-6}$ & 0.53 & 0.06 & $<1\times 10^{-6}$ \\
{\rm [Ne~V]$_{14\micron}$  -- [Ne~III]} & 0.91 & $<1\times 10^{-6}$ & 0.79 & $<1\times 10^{-6}$ & 0.75 & 0.06 & $<1\times 10^{-6}$ \\
{\rm [Ne~V]$_{14\micron}$  -- [O~IV]} & 0.94 & $<1\times 10^{-6}$ & 0.80 & $<1\times 10^{-6}$ & 0.77 & 0.06 & $<1\times 10^{-6}$ \\
{\rm [Ne~V]$_{24\micron}$ -- [Ne~V]$_{14}$} & 0.93 & $<1\times 10^{-6}$ & 0.80 & $<1\times 10^{-6}$ & 0.77 & 0.06 & $<1\times 10^{-6}$ \\
               
{\bf Seyfert 1 Galaxies} \\                
BAT -- [Ne~II] & 0.75 & $<1\times 10^{-6}$ & 0.60 & $<1\times 10^{-6}$ & 0.37 & 0.11 & $<9\times 10^{-4}$ \\
BAT -- [Ne~III] & 0.82 & $<1\times 10^{-6}$ & 0.66 & $<1\times 10^{-6}$ & 0.41 & 0.12 & $<1\times 10^{-3}$ \\
BAT -- [O~IV] & 0.72 & $<1\times 10^{-6}$ & 0.56 & $<1\times 10^{-6}$ & 0.33 & 0.11 & $<3\times 10^{-3}$ \\
BAT -- [Ne~V] & 0.61 & $<2 \times 10^{-4}$ & 0.52 & $<1\times 10^{-6}$ & 0.32 & 0.12 & $<8\times 10^{-3}$ \\
{\rm [O~IV] -- [Ne~II]} & 0.84 & $<1\times 10^{-6}$ & 0.66 & $<1\times 10^{-6}$ & 0.54 & 0.09 & $<1\times 10^{-6}$ \\
{\rm [O~IV] -- [Ne~III]}& 0.93 & $<1\times 10^{-6}$ & 0.79 & $<1\times 10^{-6}$ & 0.71 & 0.08 & $<1\times 10^{-6}$ \\
{\rm [Ne~V]$_{14\micron}$ -- [Ne~II]} & 0.77 & $<1\times 10^{-6}$ & 0.61 & $<1\times 10^{-6}$ & 0.49 & 0.08 & $<1\times 10^{-6}$ \\
{\rm [Ne~V]$_{14\micron}$ -- [Ne~III]}& 0.86 & $<1\times 10^{-6}$ & 0.75 & $<1\times 10^{-6}$ & 0.67 & 0.08 & $<1\times 10^{-6}$ \\
{\rm [Ne~V]$_{14\micron}$ -- [O~IV]}& 0.92 & $<1\times 10^{-6}$ & 0.78 & $<1\times 10^{-6}$ & 0.72 & 0.08 & $<1\times 10^{-6}$ \\
{\rm [Ne~V]$_{24\micron}$ -- [Ne~V]$_{14}$} & 0.92 & $<1\times 10^{-6}$ & 0.81 & $<1\times 10^{-6}$ & 0.77 & 0.08 & $<1\times 10^{-6}$ \\
               
{\bf  Seyfert 2 Galaxies}\\
               
BAT -- [Ne~II] & 0.63 & $<7.13\times 10^{-5}$ & 0.48 & $<7\times 10^{-5}$ & 0.40 & 0.11 & $<5\times 10^{-4}$ \\
BAT -- [Ne~III] & 0.71 & $<1\times 10^{-6}$ & 0.55 & $<1\times 10^{-6}$ & 0.45 & 0.10 & $<2\times 10^{-5}$ \\
BAT -- [O~IV] & 0.56 & $<6\times 10^{-4}$ & 0.42 & $<5\times 10^{-4}$ & 0.33 & 0.10 & $<2\times 10^{-3}$ \\
BAT -- [Ne~V] & 0.65 & $<2\times 10^{-4}$ & 0.49 & $<1\times 10^{-6}$ & 0.38 & 0.11 & $<8\times 10^{-4}$\\
{\rm [O~IV] -- [Ne~II]}& 0.76 & $<1\times 10^{-6}$ & 0.59 & $<1\times 10^{-6}$ & 0.54 & 0.09 & $<1\times 10^{-6}$ \\
{\rm [O~IV] -- [Ne~III]} & 0.94 & $<1\times 10^{-6}$ & 0.81 & $<1\times 10^{-6}$ & 0.79 & 0.08 & $<1\times 10^{-6}$ \\
{\rm [Ne~V]$_{14\micron}$ -- [Ne~II]}& 0.73 & $<4\times 10^{-5}$ & 0.58 & $<1\times 10^{-6}$ & 0.52 & 0.09 & $<1\times 10^{-6}$ \\
{\rm [Ne~V]$_{14\micron}$ -- [Ne~III]}& 0.94 & $<1\times 10^{-6}$ & 0.83 & $<1\times 10^{-6}$ & 0.80 & 0.10 & $<1\times 10^{-6}$ \\
{\rm [Ne~V]$_{14\micron}$ -- [O~IV]} & 0.94  & $<1\times 10^{-6}$ & 0.80 & $<1\times 10^{-6}$ & 0.78 & 0.09 & $<1\times 10^{-6}$ \\
{\rm [Ne~V]$_{24\micron}$ -- [Ne~V]$_{14}$} & 0.92 & $<1\times 10^{-6}$ & 0.80 & $<1\times 10^{-6}$ & 0.78 & 0.09 & $<1\times 10^{-6}$ \\

\enddata
\label{corre1}
\tablecomments{ $\rho_s$ is the Spearman rank order correlation coefficient with its associated null probability, $P_\rho$.
$\tau$ represents the generalized Kendall's correlation coefficient for censored data and $\tau_p$ is the  Kendall's  coefficient for partial correlation with censored data. $P_\tau$ and $P_\tau (p)$ are the  null probabilities for the generalized and partial Kendall's correlation test, respectively. We also present the  calculated variance, $\sigma$, for Kendall $\tau_p$. We have used a partial correlation test to exclude the effect of redshift (distance) in the Luminosity-Luminosity correlations. }
\end{deluxetable}

\clearpage

\begin{deluxetable}{lccccccc}
\tabletypesize{\scriptsize}
\tablewidth{0pt}
\tablecaption{Statistical Analysis Between Seyfert 1 and Seyfert 2 Galaxies}
\tablehead{ & \multicolumn{3}{c}{Seyfert 1}& \multicolumn{3}{c}{Seyfert 2}\\
 \cline{2-4} \cline{5-7}\\
 & \colhead{Measurements} && Standard & Measurements &&Standard &${\rm P_{K-S}}$\\
 &\colhead{Available} & Median&Deviation &Available&Median&Deviation& (\%)}
\startdata
${\rm  Log BAT/[O~IV]} $&38  &  2.5  &  0.5 & 33&  2.4     & 0.5   & 2.4    \\
${\rm Log BAT/[Ne~III]}$ &38  & 2.8 & 0.4  & 33 & 2.5   & 0.4   &  9.0   \\
${\rm  Log BAT/[Ne~V]_{14.32\micron}} $& 32  &   3.0   &  0.4   & 33 & 2.9    &  0.5 & 26.1     \\
${\rm  Log BAT/[Ne~V]_{24.32\micron}} $& 31 & 3.0    & 0.5    &30 &  2.8    & 0.3   &  9.5    \\
${\rm Log BAT/[Ne~II]}$ & 38  &  2.9   &  0.4   &  33 & 2.7   & 0.5  &  2.8 \\
${\rm [Ne~III]/[Ne~II]}$& 38    &  1.2      &  0.1       & 33    &  1.4       &   0.1    &  54.8     \\  
${\rm [O~IV]/[Ne~II]}$&  38  &    1.8     &   0.4      &  33   &    1.7     &     0.4  &   84.5    \\
${\rm [O~IV]/[Ne~III]}$&  38  &     1.5    &     0.1    &   33  &   1.7      &  0.1     &    91.4   \\
${\rm [Ne~V]_{14.32\micron}/[Ne~V]_{24.32\micron}}$&  29  &     1.0    &     0.3    &   30  &   1.0      &  0.4     &    94.0   \\
${\rm L_{BAT}}$ &  38 &  43.55   & 0.573   & 33  & 43.28   & 0.657  &  6.3  \\
${\rm L_{[Ne~II]}}$ & 38  &  40.79    & 0.613    & 33  & 40.76  & 0.799   &  82.6  \\
${\rm L_{[Ne~V]_{14.32\micron}}}$ & 32 &  40.54    &  0.659  & 33  & 40.33   & 0.886   & 69.8  \\
${\rm L_{[Ne~III]}}$ & 38  &  40.92   &  0.668  & 33  &  40.74  & 0.776   &  50.9  \\
${\rm L_{[NeV]_{24.32\micron}}}$& 31 &  40.52   &  0.623   &  30 &  40.44  & 0.746   & 28.6  \\
${\rm L_{[O~IV]}}$ & 38  &   40.98   & 0.744  & 33  & 41.00    &  0.793 & 98.1   \\

\label{table2} 
\enddata
\tablecomments{The last column, ${\rm P_{K-S}}$,  represents the Kolmogorov-Smirnov (K-S) test null probability. Upper limits for the  [Ne~V] fluxes are not included. This table also includes information about the numbers of Seyfert~1 and Seyfert~2 galaxies, median values and standard deviations of the mean for the measured quantities.
}  
\end{deluxetable}

\clearpage
\begin{deluxetable}{ccc}
\tabletypesize{\small}
\tablewidth{0pt} 
\tablecaption{Linear Regressions  for the Mid-Infrared and 14-195~keV Luminosities}
\tablehead{ \colhead{Log X - Log Y}&\colhead{Y=aX+b}\\
          &\colhead{a} &\colhead{b}}
\startdata

{\bf All Sample} \\        
BAT -- [Ne~II] & 0.72 $\pm$ 0.11 & 9.60 $\pm$ 4.85 \\
BAT -- [Ne~III] & 0.88 $\pm$ 0.06 & 2.58 $\pm$ 2.70 \\
BAT -- [O~IV] & 0.93 $\pm$ 0.08 & 0.42 $\pm$ 3.32 \\
BAT -- [Ne~V]$_{14\micron}$ & 1.00 $\pm$ 0.10 & -3.28 $\pm$ 4.51 \\
{\rm [O~IV] -- [Ne~II]} & 0.85 $\pm$ 0.07 & 5.83 $\pm$ 2.67 \\
{\rm [O~IV] -- [Ne~III]} & 0.90 $\pm$ 0.04 & 3.96 $\pm$ 1.65 \\
{\rm [Ne~III] -- [Ne~II]} & 0.94 $\pm$ 0.04 & 2.22 $\pm$ 1.80 \\
{\rm [Ne~V]$_{14\micron}$} -- {\rm [Ne~II]} & 0.67 $\pm$ 0.06 & 13.46 $\pm$ 2.35 \\
{\rm [Ne~V]$_{14\micron}$} -- {\rm [Ne~III]} & 0.82 $\pm$ 0.04 & 7.80 $\pm$ 1.43 \\
{\rm [Ne~V]$_{14\micron}$} -- {\rm [O~IV]} & 0.93 $\pm$ 0.03 & 3.59 $\pm$ 1.20 \\
{\rm [Ne~V]$_{24\micron}$} -- {\rm [Ne~V]$_{14\micron}$} & 0.89 $\pm$ 0.05 & 4.52 $\pm$ 2.20 \\

{\bf Seyfert 1 Galaxies} \\        
BAT -- [Ne~II] & 0.77 $\pm$ 0.08 & 7.08 $\pm$ 3.69 \\
BAT -- [Ne~III] & 0.93 $\pm$ 0.08 & 0.43 $\pm$ 3.68 \\
BAT -- [O~IV] & 0.93 $\pm$ 0.11 & 0.37 $\pm$ 5.00 \\
BAT -- [Ne~V]$_{14\micron}$ & 0.91 $\pm$ 0.15 & 0.79 $\pm$ 6.65 \\
{\rm [O~IV] -- [Ne~II]} & 0.84 $\pm$ 0.10 & 6.21 $\pm$ 3.91 \\
{\rm [O~IV] -- [Ne~III]} & 0.91 $\pm$ 0.07 & 3.57 $\pm$ 2.78 \\
{\rm [Ne~III] -- [Ne~II]} & 0.92 $\pm$ 0.06 & 3.03 $\pm$ 2.56 \\
{\rm [Ne~V]$_{14\micron}$} -- {\rm [Ne~II]} & 0.74 $\pm$ 0.09 & 10.65 $\pm$ 3.76 \\
{\rm [Ne~V]$_{14\micron}$} -- {\rm [Ne~III]} & 0.90 $\pm$ 0.06 & 4.44 $\pm$ 2.53 \\
{\rm [Ne~V]$_{14\micron}$} -- {\rm [O~IV]} & 1.00 $\pm$ 0.05 & 0.62 $\pm$ 2.16 \\
{\rm [Ne~V]$_{24\micron}$} -- {\rm [Ne~V]$_{14\micron}$} & 1.02 $\pm$ 0.05 & -0.69 $\pm$ 1.50 \\
         
{\bf Seyfert 2 Galaxies} \\        
BAT -- [Ne~II] & 0.94 $\pm$ 0.15 & 0.15 $\pm$ 6.56 \\
BAT -- [Ne~III] & 0.99 $\pm$ 0.10 & -1.82 $\pm$ 4.42 \\
BAT -- [O~IV] & 1.00 $\pm$ 0.13 & -2.17 $\pm$ 5.72 \\
BAT -- [Ne~V]$_{14\micron}$ & 1.10 $\pm$ 0.12 & -7.10 $\pm$ 5.28 \\
{\rm [O~IV] -- [Ne~II]} & 0.98 $\pm$ 0.10 & 0.48 $\pm$ 3.89 \\
{\rm [O~IV] -- [Ne~III]} & 0.96 $\pm$ 0.05 & 1.47 $\pm$ 2.14 \\
{\rm [Ne~III] -- [Ne~II]} & 1.02 $\pm$ 0.05 & -1.01 $\pm$ 2.18 \\
{\rm [Ne~V]$_{14\micron}$} -- {\rm [Ne~II]} & 0.89 $\pm$ 0.09 & 4.78 $\pm$ 3.62 \\
{\rm [Ne~V]$_{14\micron}$} -- {\rm [Ne~III]} & 0.87 $\pm$ 0.05 & 5.73 $\pm$ 2.11 \\
{\rm [Ne~V]$_{14\micron}$} -- {\rm [O~IV]} & 0.90 $\pm$ 0.04 & 4.45 $\pm$ 1.50 \\
{\rm [Ne~V]$_{24\micron}$} -- {\rm [Ne~V]$_{14\micron}$} & 1.07 $\pm$ 0.04 & -2.97 $\pm$ 1.48 \\

\enddata
\label{table0}
\tablecomments{$\log X$ and $\log Y$ represent the independent and dependent variables, respectively. a and b represent the regression coefficient (slope) and regression constant (intercept) respectively. For the relationship between the 14-195~keV and mid-infrared luminosities we used the ordinary least-square regression of the dependent variable, Y, against the independent variable X, OLS($Y|X$). For the relationship between the mid-infrared emission line luminosities we used the OLS bisector method  which treat the variables symmetrically \citep[see][for a review]{1990ApJ...364..104I}. When censored data was present we used the EM and the Buckley-James (BJ) method, if censored data was present in both variables (e.g.,  the  ${\rm [Ne~V]_{14}-[Ne V]_{24}}$ relationship) the Schmitt's binned method was used \citep{1986ApJ...306..490I,1990BAAS...22..917I}. The results from the EM and BJ methods agree within their estimated errors, therefore, for the sake of simplicity, we present only the values from the EM regression.}
\end{deluxetable}

\end{document}